\documentclass[journal]{IEEEtran}

\usepackage{epsfig,latexsym}
\usepackage{float}
\usepackage{indentfirst}
\usepackage{amsmath}
\usepackage{amssymb}
\usepackage{times}
\usepackage{psfrag}
\usepackage{cite}
\usepackage{subfigure}
\usepackage{textcomp}
\usepackage{multirow}
\usepackage{multicol}
\usepackage{stfloats}
\usepackage{url}
\usepackage{subfigure}
\usepackage{multirow}
\usepackage{booktabs}
\usepackage{graphicx}
\usepackage{bm}
\usepackage{color}
\usepackage{mathrsfs}

\graphicspath{
    {./images/}{./images/Heatmaps/}{./images/V-DMOS/}{./images/Objective/}{./images/QP/}
}

\newcommand{\tabincell}[2]{\begin{tabular}{@{}#1@{}}#2\end{tabular}}

\newcommand{\argmax}{\operatornamewithlimits{argmax}}

\begin{document}
\title{Assessing Visual Quality of Omnidirectional Videos}
\author{Mai~Xu,~\IEEEmembership{Senior Member,~IEEE,} Chen~Li,
Zhenzhong~Chen~\IEEEmembership{Senior Member,~IEEE}, Zulin~Wang,~\IEEEmembership{Member,~IEEE} and Zhenyu Guan,~\IEEEmembership{Member,~IEEE}\\
\thanks{M. Xu, C. Li, Z. Wang and Z. Guan are with the School of Electronic and Information Engineering, Beihang University, Beijing, 100191 China (e-mail: Maixu@buaa.edu.cn; jnlichen123@buaa.edu.cn; wzulin@buaa.edu.cn; guanzhenyu@buaa.edu.cn). Z. Chen is with Wuhan University, Wuhan, China (e-mail: zzchen@whu.edu.cn). This work was supported by NSFC under grant number 61573037.}}
\maketitle
\begin{abstract}
In contrast with traditional video, omnidirectional video enables spherical viewing direction with support for head-mounted displays, providing an interactive and immersive experience. Unfortunately, to the best of our knowledge, there are few visual quality assessment (VQA) methods, either subjective or objective, for omnidirectional video coding. This paper proposes both subjective and objective methods for assessing quality loss in encoding omnidirectional video. Specifically, we first present a new database, which includes the viewing direction data from several subjects watching omnidirectional video sequences. Then, from our database, we find a high consistency in viewing directions across different subjects. The viewing directions are normally distributed in the center of the front regions, but they sometimes fall into other regions, related to video content.
Given this finding, we present a subjective VQA method for measuring difference mean opinion score (DMOS) of the whole and regional omnidirectional video, in terms of overall DMOS (O-DMOS) and vectorized DMOS (V-DMOS), respectively.
Moreover, we propose two objective VQA methods for encoded omnidirectional video, in light of human perception characteristics of omnidirectional video. One method weighs the distortion of pixels with regard to their distances to the center of front regions, which considers human preference in a panorama.
The other method predicts viewing directions according to video content, and then the predicted viewing directions are leveraged to allocate weights to the distortion of each pixel in our objective VQA method.
Finally, our experimental results verify that both the subjective and objective methods proposed in this paper advance state-of-the-art VQA for omnidirectional video.
\end{abstract}
\section{Introduction}\label{intro}
Recent years have witnessed the rapid development of virtual reality (VR). According to a report by \cite{ilabreport}, $90\%$ VR content is in the form of omnidirectional video, which involves a $360^{\circ}\times180^{\circ}$ viewing direction. With the support of head-mounted displays (HMD), omnidirectional video offers an immersive and even interactive experience~\cite{sarmiento2009panoramic}. On the other hand, it is likely that the quality of experience (QoE)~\cite{konrad2016novel} of omnidirectional video dramatically degrades when presented at low resolutions or with compression artifacts. Such QoE degradation always makes humans feel uncomfortable, as reported in the MPEG survey \cite{mpeg2016vrsurvey}. Therefore, it is necessary to study visual quality assessment (VQA) for omnidirectional video coding.

Both subjective and objective methods are needed for Full-Reference (FR) VQA in omnidirectional video coding \cite{FuRhombic2009}. Subjective VQA refers to measuring the quality of omnidirectional video as rated by humans. Since omnidirectional video ultimately outputs to human eyes, subjective VQA is more rational than objective VQA in assessing visual quality. In addition, subjective VQA can be also used to verify the effectiveness of objective VQA. There are several subjective VQA methods \cite{upenik2016testbed, zakharchenko2016quality,schatz2017towards,singla2017measuring,singla2017comparison} for measuring the quality reduction of encoded omnidirectional videos, and most of them simply apply score processing metrics for traditional 2D videos. In contrast, there are a great number of subjective VQA methods for traditional 2D videos, such as \cite{itu1998710, rec2008p, rec2012bt}. In these methods, the mean opinion score (MOS) \cite{tan2016video} and difference MOS (DMOS) \cite{seshadrinathan2010subjective} are widely used metrics for subjective VQA. In this paper, we thus propose a subjective VQA method to assess quality loss in encoding omnidirectional video in the form of DMOS. Different from the latest subjective VQA methods of omnidirectional video \cite{upenik2016testbed, zakharchenko2016quality,schatz2017towards,singla2017measuring,singla2017comparison}, our method proposes the least subject number and a new test procedure according to our findings on human behavior of viewing omnidirectional video. In addition, two metrics of DMOS, i.e., the overall DMOS (O-DMOS) and vectorized DMOS (V-DMOS), are proposed in our subjective VQA method.

For objective VQA, the spherical characteristic of omnidirectional video has been taken into account in the latest work of \cite{zakharchenko2016quality, yu2015framework}. For example, Yu \textit{et al.}~\cite{yu2015framework} proposed a sphere-based peak signal-to-noise ratio (S-PSNR), which calculates peak signal-to-noise ratio (PSNR) based on a set of uniformly sampled points on a sphere instead of rectangularly mapped pixels. By applying interpolation algorithms, S-PSNR is able to cope with objective quality assessment for omnidirectional video coding under different projections. The main difference between 2D and omnidirectional videos is that only content inside the field of view (FoV) is accessible in omnidirectional video. However, none of the existing VQA methods takes into consideration such perceptual characteristics of omnidirectional video. In this paper, we further propose to objectively assess the perceptual quality of encoded omnidirectional video by considering the accessible FoV of possible viewing directions.
To the best of our knowledge, our work is the first attempt to predict possible FoV in weighting distortion of omnidirectional video, despite the weight mechanism being widely used in VQA \cite{yu2015framework, lee2002foveated, LiClosed2017}.

To be more specific, this paper first presents a new database containing the viewing directions of 40 subjects watching 48 omnidirectional video sequences. Then, by mining our database, we discover that the viewing directions across different subjects are highly consistent. In light of such a finding, we develop two subjective VQA metrics, namely O-DMOS and V-DMOS, for rating the overall and regional visual quality reduction of encoded omnidirectional video, respectively. We further find from our database that the consistent viewing directions of humans are related to both spherical location and video content. Accordingly, we propose two objective VQA methods that allocate weights to the distortion of each pixel when calculating the PSNR. The weight allocation in the first method only leverages humans' preferences for the location of pixels in omnidirectional video, while the second method also depends on video content.
There are some works on projection \cite{chen2018recent} and coding optimization \cite{tang2017optimized,li2017projection,he2017motion} for omnidirectional video, which emphasize region of interest (RoI). Similarly, our methods also consider RoI in VQA. V-DMOS is able to represent the quality of different regions and guide bit allocation in coding optimization.
For objective VQA methods, different regions are unequally weighted according to the distribution of viewing directions in calculating NCP-PSNR. In the CP-PSNR method, the viewing directions are further predicted, such that the quality of omnidirectional video is calculated according to RoI.

This paper is an extended version of our conference paper \cite{subjective2017icme}. Beyond the subjective VQA method in \cite{subjective2017icme}, this paper further proposes two objective VQA methods for measuring the quality of omnidirectional video coding. For the objective VQA methods, this paper also investigates some new findings about the distribution of human viewing direction in omnidirectional video. Our contributions in this paper are three-fold:
\begin{itemize}
  \item We present a viewing direction database for omnidirectional video, with a consistency analysis on the viewing directions of different subjects. We also present a VQA database consisting of raw and compressed omnidirectional sequences, in which both the subjective scores and viewing directions of different subjects are collected.
  \item We develop a new method for the subjective VQA of omnidirectional video, taking advantage of the analysis of viewing directions over our database. Our subjective VQA method has been adopted \cite{subjectiveavs} by the international standard IEEE 1857.9/AVS-VR.
  \item We propose two methods for the objective VQA of omnidirectional video coding, taking into account human perception related to spherical location and video content, respectively. Our objective VQA methods are a pioneering attempt that embeds human perception in assessing visual quality of omnidirectional video coding.
\end{itemize}
\section{Related work}
\subsection{Related work on subjective VQA}
The past two decades have witnessed a number of subjective VQA methods for 2D video. In particular, the international telecommunication union (ITU) has proposed several subjective methodologies \cite{itu1998710, rec2008p, rec2012bt} for assessing 2D video. Among these proposals, the double stimulus continuous quality scale (DSCQS) \cite{pinson2003comparing}, single stimulus continuous quality scale (SSCQS) \cite{seshadrinathan2010subjective} and single stimulus continuous quality evaluation (SSCQE) \cite{lee2006comparison} were adopted to determine the display orders of sequences when viewing and rating video sequences. Additionally, two metrics have been widely used in rating the subjective VQA of 2D video: one metric is MOS \cite{tan2016video} for no-reference (NR), reduced-reference (RR) and FR assessments; the other metric is DMOS \cite{seshadrinathan2010subjective, seshadrinathan2010study}, which is for FR assessment only. Recently, several subjective VQA methods for other types of videos have emerged. For example, Pourashraf \textit{et al.} \cite{pourashraf2014minimisation} proposed measuring the subjective quality of video conferencing by adopting DMOS for the conventional subjective VQA method. ITU extended their DMOS-based VQA method for stereoscopic video~\cite{union2015subjective}, which incorporates the characteristics of stereoscopic video.

Although omnidirectional video is flooding into our daily life, there are some works in the literature \cite{upenik2016testbed, zakharchenko2016quality,schatz2017towards,singla2017measuring,singla2017comparison} on the subjective VQA of omnidirectional video. Upenik \textit{et al.} \cite{upenik2016testbed} proposed a testbed for subjective VQA on omnidirectional video and image. In their testbed, an HMD is suggested as the displaying device, and a custom software application is provided. Unfortunately, \cite{upenik2016testbed} does not deal with how to measure the subjective quality of omnidirectional video coding.
Zakharchenko \textit{et al.} conducted the subjective VQA experiment to validate their objective VQA methods for omnidirectional video \cite{zakharchenko2016quality}. In \cite{zakharchenko2016quality}, subjects were forced to view one region of omnidirectional video, and then the conventional subjective VQA method for 2D video is simply applied. However, this is not in accordance with the interactive experience on omnidirectional video. More importantly, an immersive experience cannot be achieved in the subjective VQA of \cite{zakharchenko2016quality}, such that the DMOS does not meet practical QoE for humans.
Schatz \textit{et al.} developed an approach towards QoE of omnidirectional video streaming \cite{schatz2017towards}, which mainly focuses on the impact of stalling on omnidirectional streaming.
Singla \textit{et al.} proposed a modified absolute category rating (M-ACR) method, analyzing subjective quality and simulator sickness of omnidirectional videos at varying resolutions and bit-rates across different devices \cite{singla2017measuring,singla2017comparison}.
However, none of the above methods considers human behavior and interactive experience in subjective VQA, which can be mined from the viewing direction data of many subjects.
In this paper, we propose a subjective VQA method that considers the interactive behavior of humans in viewing omnidirectional video, such that the QoE of subjects can be reflected in our subjective metric.
\subsection{Related work on objective VQA}
For objective VQA of 2D video, a commonly used FR metric is PSNR. PSNR is based on the mean squared error (MSE) between the reference and processed videos and has been well
studied from a mathematical perspective. However, PSNR cannot successfully reflect the subjective visual quality perceived by the human visual system (HVS), as it does not consider human perception at all.
For example, the subjective quality is more likely to be influenced by the PSNR in RoI.
In order to better correlate assessments with subjective quality, many advanced PNSR-based methods \cite{lee2002foveated, chou1996perceptually, cavallaro2005semantic, li2011visual, chen2010perceptually, Xu2014JSTP, liu2017free} have been proposed to improve the existing PSNR metric for the VQA of 2D video by accounting for the importance of each pixel. For example, based on the foveation response of HVS, foveal PSNR (FPSNR) \cite{lee2002foveated} was proposed, using a non-uniform resolution weighting metric, in which the distortion weights decrease with eccentricity.
The peak signal-to-perceptible noise ratio (PSPNR) \cite{chou1996perceptually} and foveated PSPNR \cite{chen2010perceptually} were presented to consider distortion, only when the errors are larger than the just-noticeable-distortion (JND) thresholds. Similarly, semantic PSNR (SPSNR) \cite{cavallaro2005semantic} and eye-tracking-weighted PSNR (EWPSNR) \cite{li2011visual} were developed based on the form of PSNR as well. EWPSNR has a better performance in evaluating visual quality according to the real-time detected eye fixation points. In \cite{Xu2014JSTP}, a weight-based PSNR metric was proposed to measure the quality of video conferencing. Their method imposes a greater penalty weight on regions with faces and facial features when calculating the PSNR. Free energy adjusted PSNR (FEA-PSNR) was proposed in \cite{liu2017free}. This method considers image perceptual complexity when assessing image quality. In \cite{na2014novel}, a no-reference PSNR method was proposed to assess both quantization error and the blocky effect in measuring the non-reference quality of H.264/AVC videos.

In addition to these PSNR-based methods, there are many other objective VQA methods for 2D video. For example, the universal image quality index (UQI) \cite{wang2002universal} and its improved form, the single-scale structural similarity index (SSIM) \cite{wang2004image}, were proposed to model the image distortion as the combination of losses on luminance, contrast, and structure. The information fidelity criterion (IFC) \cite{sheikh2005information} is based on information-theory, in which the visual quality is qualified as a mutual information between the impaired and reference images. The visual signal to noise ratio (VSNR) \cite{chandler2007vsnr} is a wavelet based method, which detects the visibility of distortions.

For omnidirectional video, there are several objective VQA works \cite{zakharchenko2016quality, yu2015framework, upenik2017performance}, also based on PSNR. In evaluating the quality degradation of omnidirectional video encoding, the work of \cite{zakharchenko2016quality, yu2015framework} takes into account the spherical characteristic of omnidirectional video. For example, Yu \textit{et al.} \cite{yu2015framework} proposed S-PSNR, which calculates PSNR based on a set of uniformly sampled points on a sphere instead of rectangularly mapped pixels. By applying interpolation algorithms, S-PSNR is able to generate objective quality assessments for encoded omnidirectional videos under different projections.
Additionally, Zakharchenkoa \textit{et al.} \cite{zakharchenko2016quality} proposed a weighted PSNR (W-PSNR) using gamma-corrected pixel values for the PSNR calculation process. Craster parabolic projection PSNR (CPP-PSNR) was also proposed to convert the projection of omnidirectional video to Craster parabolic projection (CPP) for calculating PSNR.
The latest work of \cite{upenik2017performance} conducted an experiment to evaluate the performance of several objective VQA methods on encoded omnidirectional images, via measuring the correlation between objective and subjective quality. The experimental results reveal that the VQA methods designed for omnidirectional content slightly outperform traditional VQA methods for 2D content.
This finding is probably because none of the existing VQA methods explores the human perception model for omnidirectional video, in which only content inside FoV is accessible. Therefore, this paper further proposes to objectively assess visual quality of omnidirectional video coding, by considering the FoV of possible viewing directions.

\subsection{Related work on database of omnidirectional content}
Since our VQA methods rely on the analysis of human behavior in viewing omnidirectional video, it is necessary to  build a database consisting of human behavior data in omnidirectional content.
For omnidirectional images, there exist several works \cite{rai2017saliency, de2017look} on collecting viewing direction and eye gaze data.
For omnidirectional video, some databases were also built including the viewing direction data.
Specifically, Corbillon \textit{et al.} introduced a database \cite{corbillon2017360} with the viewing direction data in omnidirectional video sequences. However, the number of omnidirectional sequences is only 7, too small for a thorough analysis of human behavior.
Wu \textit{et al.} \cite{wu2017dataset} presented a larger database, which includes 18 omnidirectional video sequences. The sequences in \cite{wu2017dataset} are still insufficient with non-diverse content for thoroughly analyzing the viewing directions of humans in omnidirectional video.
Therefore, we establish a new database that is composed of 48 omnidirectional video sequences with diverse content, in order to analyze viewing directions of different subjects. 
Based on the newly established database, we report some findings through the analysis of viewing directions across subjects.
Our analysis mainly refers to human consistency in viewing omnidirectional video,
which has not been discovered in existing works \cite{singla2017measuring,singla2017comparison} on user behavior analysis of omnidirectional video.
We also established a VQA database for omnidirectional video that consists of 12 raw sequences and 36 compressed sequences. With this database, fair performance evaluation among different VQA methods can be implemented.
%
\begin{table*}[!tb]
\caption{Omnidirectional video test sequence categories.} \label{tab:video-category}
\vspace{-3em}
\begin{center}
\resizebox{\textwidth}{!}{
\begin{tabular}{*{10}{|c}|}
  \hline
  \textbf{Category} & Computer Animation (CA) & Driving & Action Sports & Movie & Video Game & Scenery & Show & Others & \textbf{In Total} \\
  \hline
  \textbf{Number of Video sequences} & 6 & 6 & 6 & 6 & 6 & 6 & 6 & 6 & 48 \\
  \hline
\end{tabular}
}
\end{center}
\vspace{-2em}
\end{table*}
\section{Analysis of consistency in viewing omnidirectional video}
Due to its omnidirectionality, people cannot see the whole content of omnidirectional video from a single viewing position for instance. Instead, they normally look around and focus on what attracts them. It is intuitive that there may exist consistency across different subjects in their viewing directions on watching omnidirectional video. Thus, this section mainly discusses the analysis of consistency in viewing omnidirectional video.
\subsection{Database}
\label{sec:database}
We present a new database that contains viewing direction data from 40 subjects when watching omnidirectional video sequences. In all, there are 48 sequences of omnidirectional video in our database.
To ensure good QoE, the resolution of the sequences is chosen beyond 3K ($2880\times1440$) and up to 8K ($7680\times3840$).
These sequences are diverse in terms of their content, and they can be categorized according to video content, as shown in Table~\ref{tab:video-category}. All of these 48 sequences were downloaded from YouTube or VRCun. Then, the sequences were cut into short clips with durations ranging from 20 to 60 seconds.
Following \cite{seshadrinathan2010subjective,seshadrinathan2010study}, the audio tracks were discarded to avoid the impacts of acoustic information.

We used the HTC Vive as the HMD and a software Virtual Desktop (VD) as the omnidirectional video player. In total, 40 subjects (29 male and 11 female) participated in the experiment. For each subject, all of the 48 sequences were displayed in a random order. During the experiment, the subjects were seated in a swivel chair and were allowed to turn around freely, such that all regions of omnidirectional video were accessible.
Although the subjects are able to move horizontally with the swivel chair, the viewport is only determined by the viewing direction instead of the horizontal moving.
Besides, to avoid eye fatigue and motion sickness \cite{mpeg2016vrsurvey}, there was a 5-minute interval between every 16-sequence session. With the support of the Vive software development kit (SDK), we were able to collect the posture data of subjects when viewing omnidirectional video. Then, the viewing direction data describing where subjects paid attention were obtained in the form of Euler angles, and only the inclination and azimuth angles were recorded. Based on the inclination and azimuth angles, viewing directions of each subject, in terms of longitude and latitude, were collected for the omnidirectional video sequences in our database. Our database is available at \url{https://github.com/Archer-Tatsu/head-tracking}.
\subsection{Data analysis}
\label{sec:data-analysis}
We now analyze the viewing direction data in our database. First, we discard the viewing direction data of the first second in each sequence since the viewing directions of all subjects were initialized to be in the center of the front region. The remaining data are then used for our analysis. Our findings with the corresponding analysis are presented and analyzed as follows.
\begin{table*}[!tb]
\begin{center}
  \caption{CC of viewing direction heat maps between Groups $A$ and $B$ for each omnidirectional video sequence} \label{tab:CC}
  \vspace{-2em}
  \resizebox{\textwidth}{!}{
  \begin{tabular}{*{4}{|c|c|c}|}
  \hline
  \tabincell{c}{Cate-\\gory} & Name & CC & \tabincell{c}{Cate-\\gory} & Name & CC & \tabincell{c}{Cate-\\gory} & Name & CC & \tabincell{c}{Cate-\\gory} & Name & CC \\
  \hline
  \multirow{6}{*}{\rotatebox{90}{CA}} & AcerPredator & 0.839$\pm$0.087
&
  \multirow{6}{*}{\rotatebox{90}{Driving}} & AirShow & 0.783$\pm$0.078
 &
  \multirow{6}{*}{\rotatebox{90}{Others}} & A380 & 0.839$\pm$0.106
 &
  \multirow{6}{*}{\rotatebox{90}{Video Game}} & CS & 0.819$\pm$0.084
 \\
  \cline{2-3} \cline{5-6} \cline{8-9} \cline{11-12}
  & BFG & 0.644$\pm$0.146
 & & DrivingInAlps & 0.857$\pm$0.071
 & & CandyCarnival & 0.723$\pm$0.094
 & & Dota2 & 0.714$\pm$0.103
 \\
  \cline{2-3} \cline{5-6} \cline{8-9} \cline{11-12}
  & CMLauncher & 0.828$\pm$0.119
 & & F5Fighter & 0.592$\pm$0.126
 & & MercedesBenz & 0.592$\pm$0.133
 & & GalaxyOnFire & 0.762$\pm$0.084
 \\
  \cline{2-3} \cline{5-6} \cline{8-9} \cline{11-12}
  & Cryogenian & 0.526$\pm$0.174
 & & HondaF1 & 0.872$\pm$0.053
 & & RingMan & 0.897$\pm$0.054
 & & LOL & 0.724$\pm$0.097
 \\
  \cline{2-3} \cline{5-6} \cline{8-9} \cline{11-12}
  & LoopUniverse & 0.779$\pm$0.078
 & & Rally & 0.867$\pm$0.047
 & & RioOlympics & 0.624$\pm$0.123
 & & MC & 0.726$\pm$0.115
 \\
  \cline{2-3} \cline{5-6} \cline{8-9} \cline{11-12}
  & Pokemon & 0.607$\pm$0.182
 & & Supercar & 0.854$\pm$0.064
 & & VRBasketball & 0.770$\pm$0.105
 & & SuperMario64 & 0.860$\pm$0.054
 \\
  \hline
  \multirow{6}{*}{\rotatebox{90}{Movie}} & Help & 0.859$\pm$0.122
 &
  \multirow{6}{*}{\rotatebox{90}{Scenery}} & Antarctic & 0.674$\pm$0.135
 &
  \multirow{6}{*}{\rotatebox{90}{Show}} & BTSRun & 0.867$\pm$0.061
 &
  \multirow{6}{*}{\rotatebox{90}{Action Sports}} & Gliding & 0.528$\pm$0.158
 \\
  \cline{2-3} \cline{5-6} \cline{8-9} \cline{11-12}
  & IRobot & 0.771$\pm$0.078
 & & BlueWorld & 0.559$\pm$0.156
 & & Graffiti & 0.807$\pm$0.100
 & & Parachuting & 0.628$\pm$0.157
 \\
  \cline{2-3} \cline{5-6} \cline{8-9} \cline{11-12}
  & Predator & 0.696$\pm$0.124
 & & Dubai & 0.646$\pm$0.133
 & & KasabianLive & 0.722$\pm$0.132
 & & RollerCoaster & 0.834$\pm$0.078
 \\
  \cline{2-3} \cline{5-6} \cline{8-9} \cline{11-12}
  & ProjectSoul & 0.918$\pm$0.053
 & & Egypt & 0.665$\pm$0.131
 & & NotBeAloneTonight & 0.587$\pm$0.131
 & & Skiing & 0.766$\pm$0.104
 \\
  \cline{2-3} \cline{5-6} \cline{8-9} \cline{11-12}
  & StarWars & 0.950$\pm$0.016
 & & StarryPolar & 0.495$\pm$0.152
 & & Symphony & 0.779$\pm$0.096
 & & Surfing & 0.830$\pm$0.096
 \\
  \cline{2-3} \cline{5-6} \cline{8-9} \cline{11-12}
  & Terminator & 0.843$\pm$0.078
 & & WesternSichuan & 0.667$\pm$0.138
 & & VRBasketball & 0.770$\pm$0.105
 & & Waterskiing & 0.781$\pm$0.128
 \\
  \hline
  \multicolumn{2}{|c|}{Overall} & 0.745$\pm$ 0.114 & \multicolumn{9}{c|}{}\\
  \hline
  \end{tabular}}
\end{center}
\vspace{-2em}
\end{table*}
\begin{figure*}[!tb]
\begin{center}
\resizebox{\textwidth}{!}{
\hspace{-0.07in}
\subfigure[HondaF1]{
  \label{fig:heatmap:a}
  \includegraphics[width=0.22\textwidth]{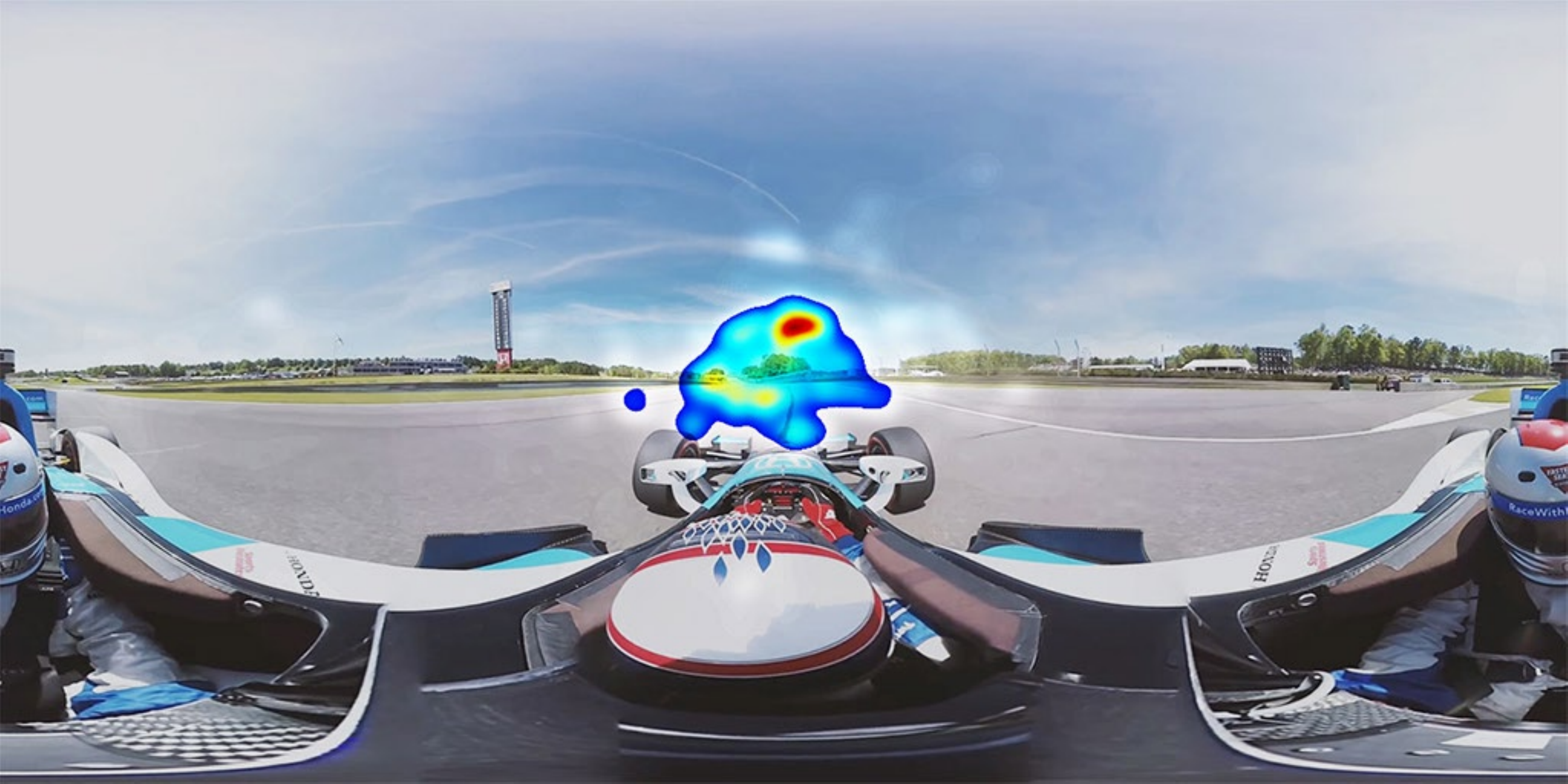}
}
\hspace{-0.1in}
\subfigure[RingMan]{
  \label{fig:heatmap:b}
  \includegraphics[width=0.22\textwidth]{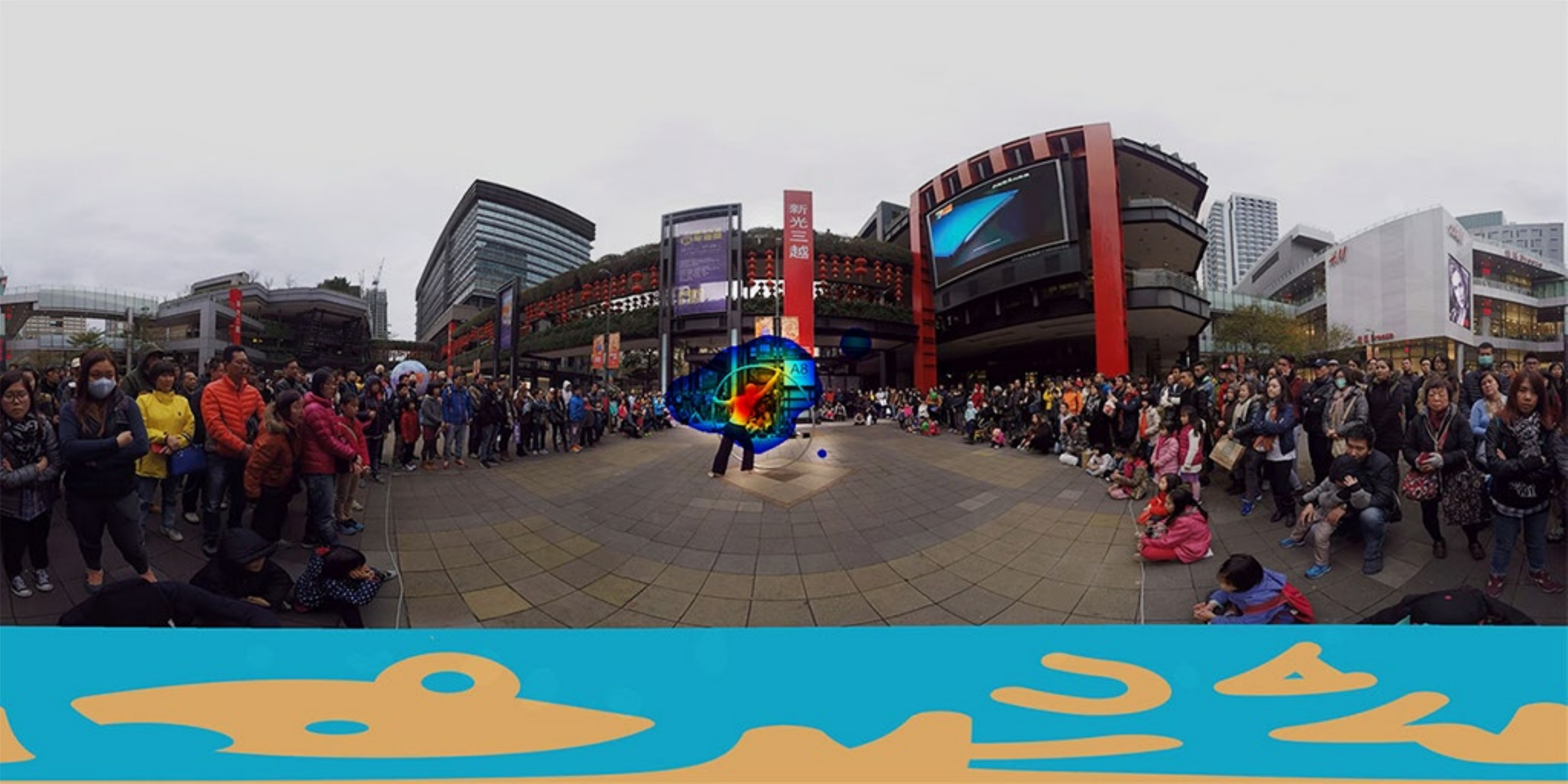}
}
\hspace{-0.1in}
\subfigure[RollerCoaster]{
  \label{fig:heatmap:c}
  \includegraphics[width=0.22\textwidth]{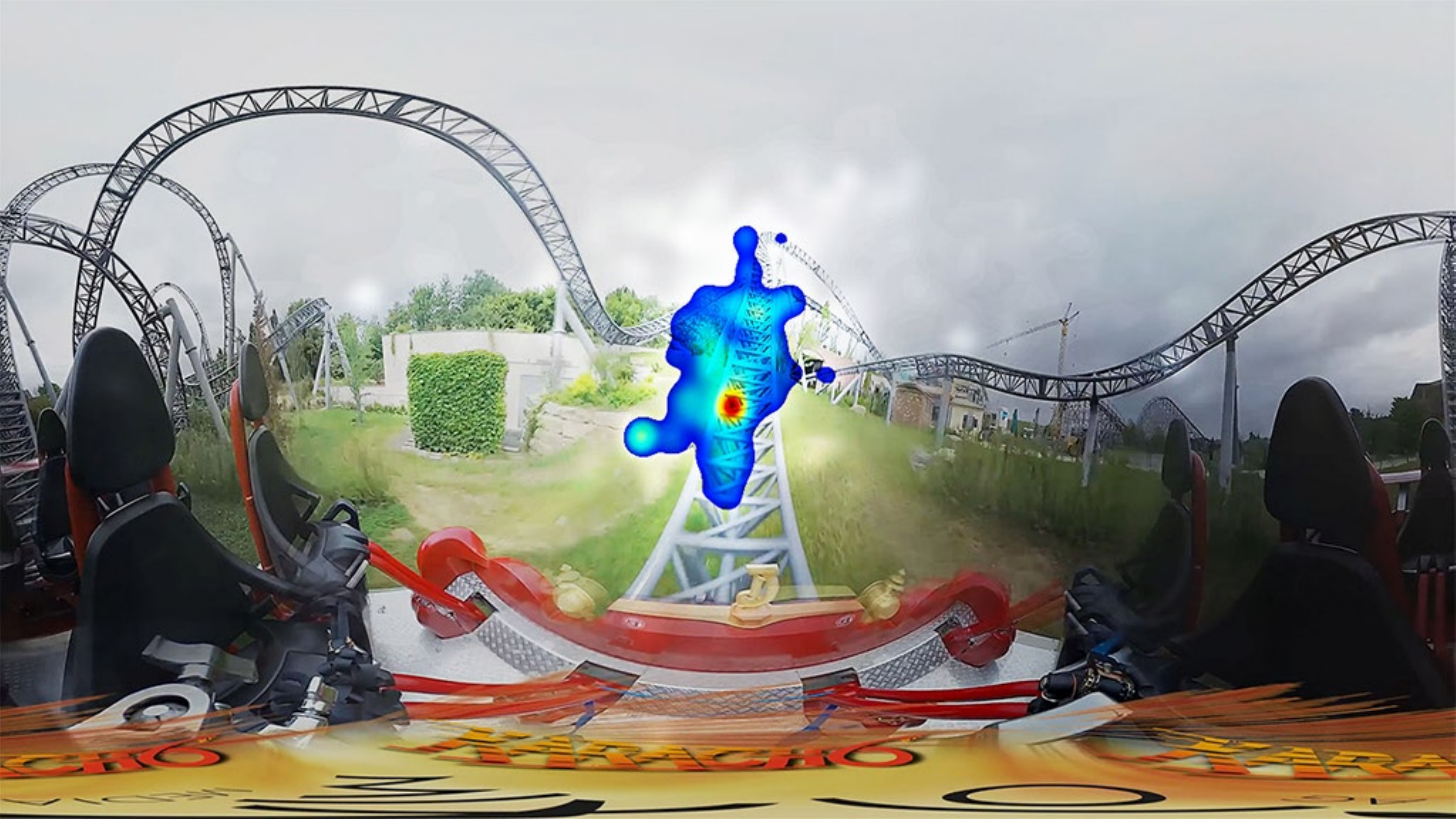}
}
\hspace{-0.1in}
\subfigure[StarWars]{
  \label{fig:heatmap:d}
  \includegraphics[width=0.22\textwidth]{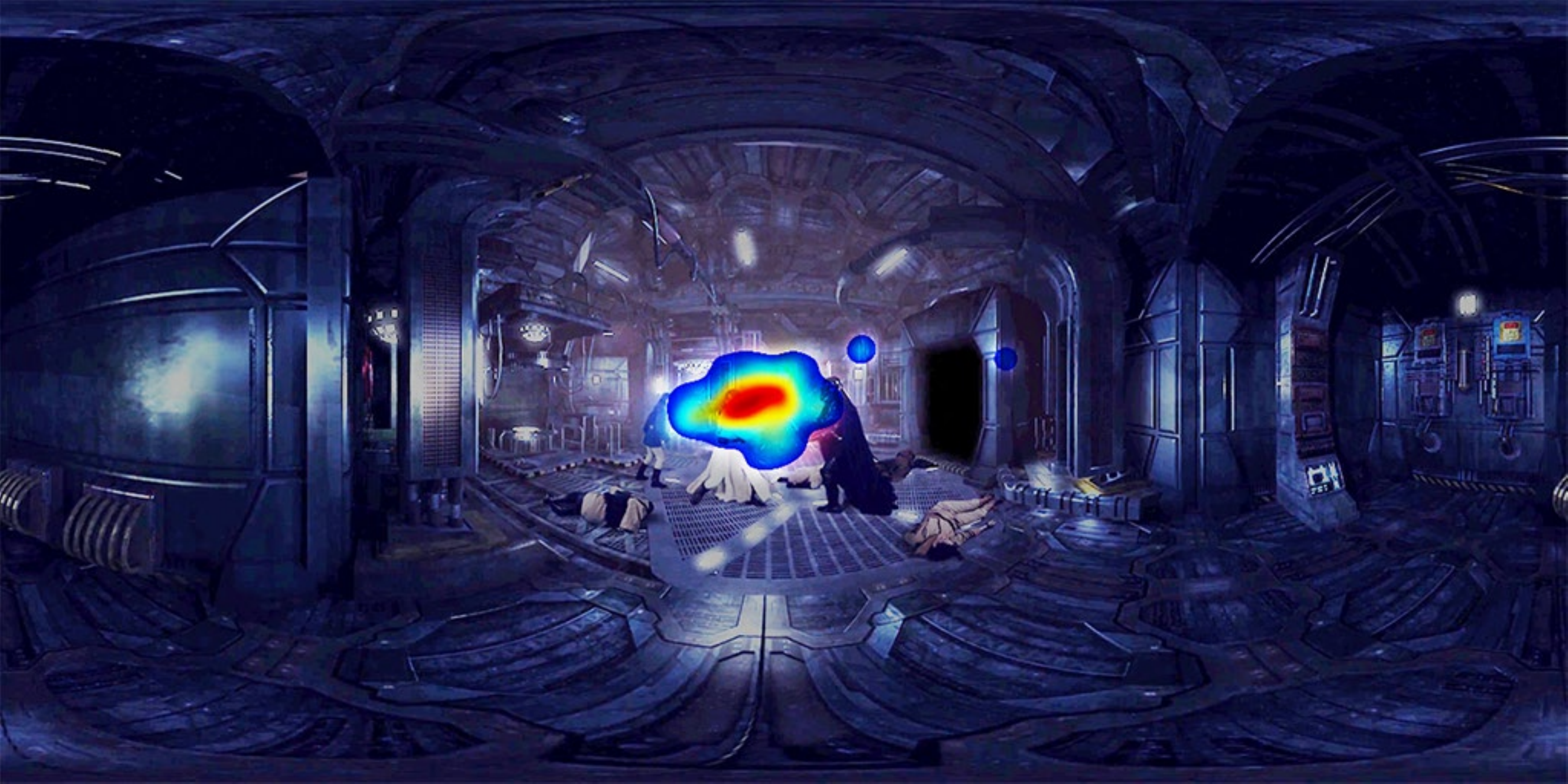}
}
\hspace{-0.1in}
\subfigure[Symphony]{
  \label{fig:heatmap:e}
  \includegraphics[width=0.22\textwidth]{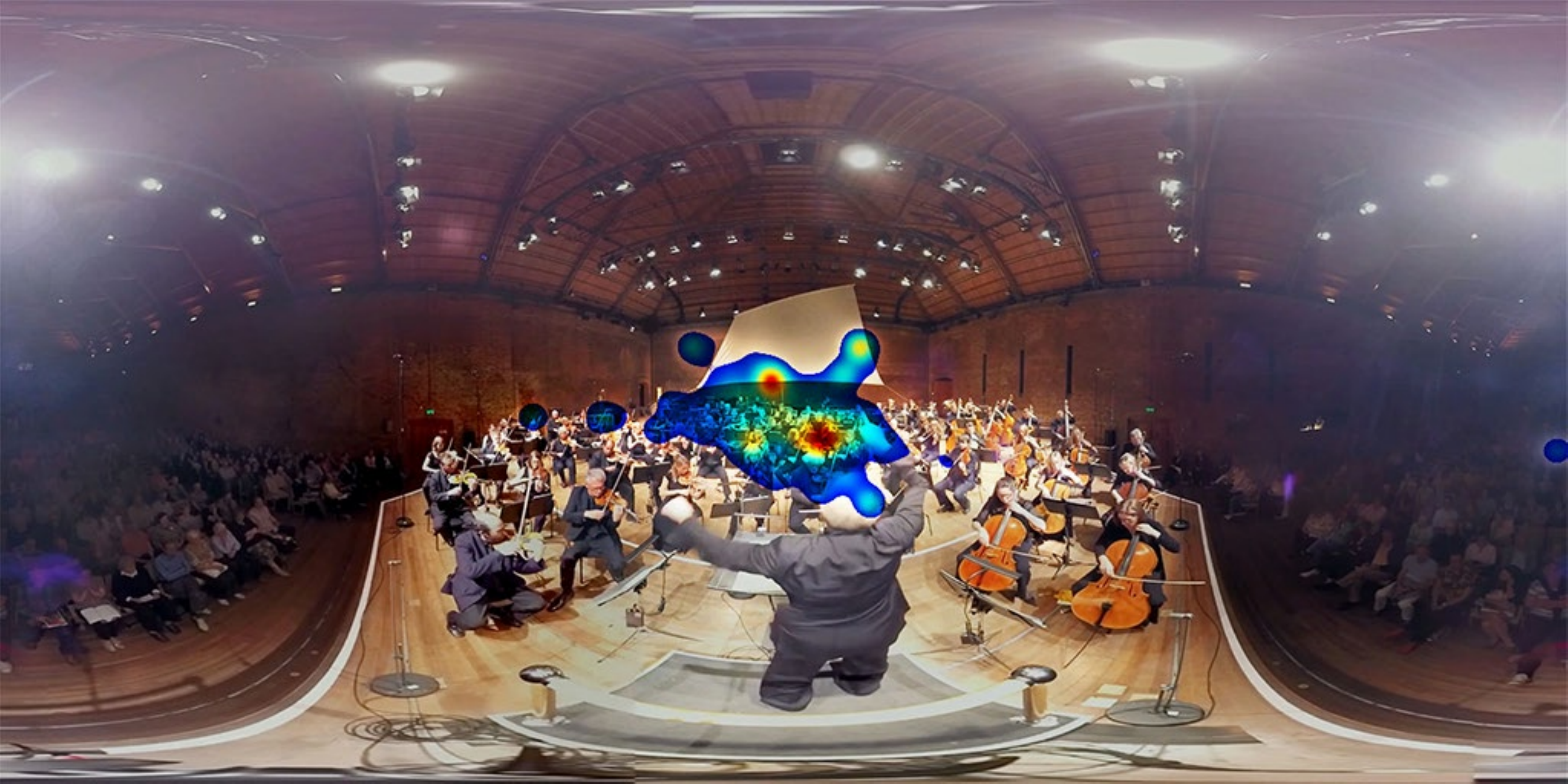}
}
\hspace{-0.07in}
}
\end{center}
\vspace{-1.3em}
\caption{Heat maps of viewing directions on some selected sequences. Note that the heat maps are obtained via the Gaussian convolution of viewing direction data for all frames viewed by 40 subjects, and the results are shown together with one randomly selected frame from each sequence.}
\label{fig:heatmap}
\vspace{-0.8em}
\end{figure*}

\textit{Finding 1: When subjects are watching omnidirectional video, the longitude and latitude of their viewing directions are almost uncorrelated with each other.}

The viewing direction data we collected in Section \ref{sec:database} consist of two dimensions, i.e., the longitude and latitude in a spherical coordinate system. Let $\bm{\varphi}$ and $\bm{\theta}$ denote the collections of the longitude and latitude of viewing directions, respectively, from all omnidirectional video sequences in our database. Then, the covariance between $\bm{\varphi}$ and $\bm{\theta}$ can be calculated as follows,
\begin{equation}
\label{eq:cov}
\small
\mathrm{cov}(\bm{\varphi},\bm{\theta}) = \mathrm{E}[(\bm{\varphi}-\mathrm{E}(\bm{\varphi}))(\bm{\theta}-\mathrm{E}(\bm{\theta}))]
\mbox{,}
\end{equation}
where $\mathrm{E}(\cdot)$ is the expectation operator.
Given the covariance of \eqref{eq:cov}, the correlation between the longitude and latitude of viewing direction can be computed by
\begin{equation}
\small
\rho(\bm{\varphi},\bm{\theta}) = \frac{\mathrm{cov}(\bm{\varphi},\bm{\theta})}{\sqrt{\mathrm{var}(\bm{\varphi})}\sqrt{\mathrm{var}(\bm{\theta})}}
\mbox{,}
\end{equation}
where $\mathrm{var}(\bm{\varphi})$ and $\mathrm{var}(\bm{\theta})$ are the variances of $\bm{\varphi}$ and $\bm{\theta}$, respectively. In our database, we have $\rho(\bm{\varphi},\bm{\theta}) = -0.0337$ for the viewing directions of all omnidirectional video sequences. Since such a value of $\rho(\bm{\varphi},\bm{\theta})$ is close to 0, the correlation between the longitude and latitude of human viewing directions in omnidirectional video is rather small. This completes the analysis of \textit{Finding 1}.
\begin{figure}[!tb]
\centering
\resizebox{\linewidth}{!}{
\hspace{-0.07in}
\subfigure{
  \label{fig:frequency:a}
  \includegraphics[width=0.4\linewidth]{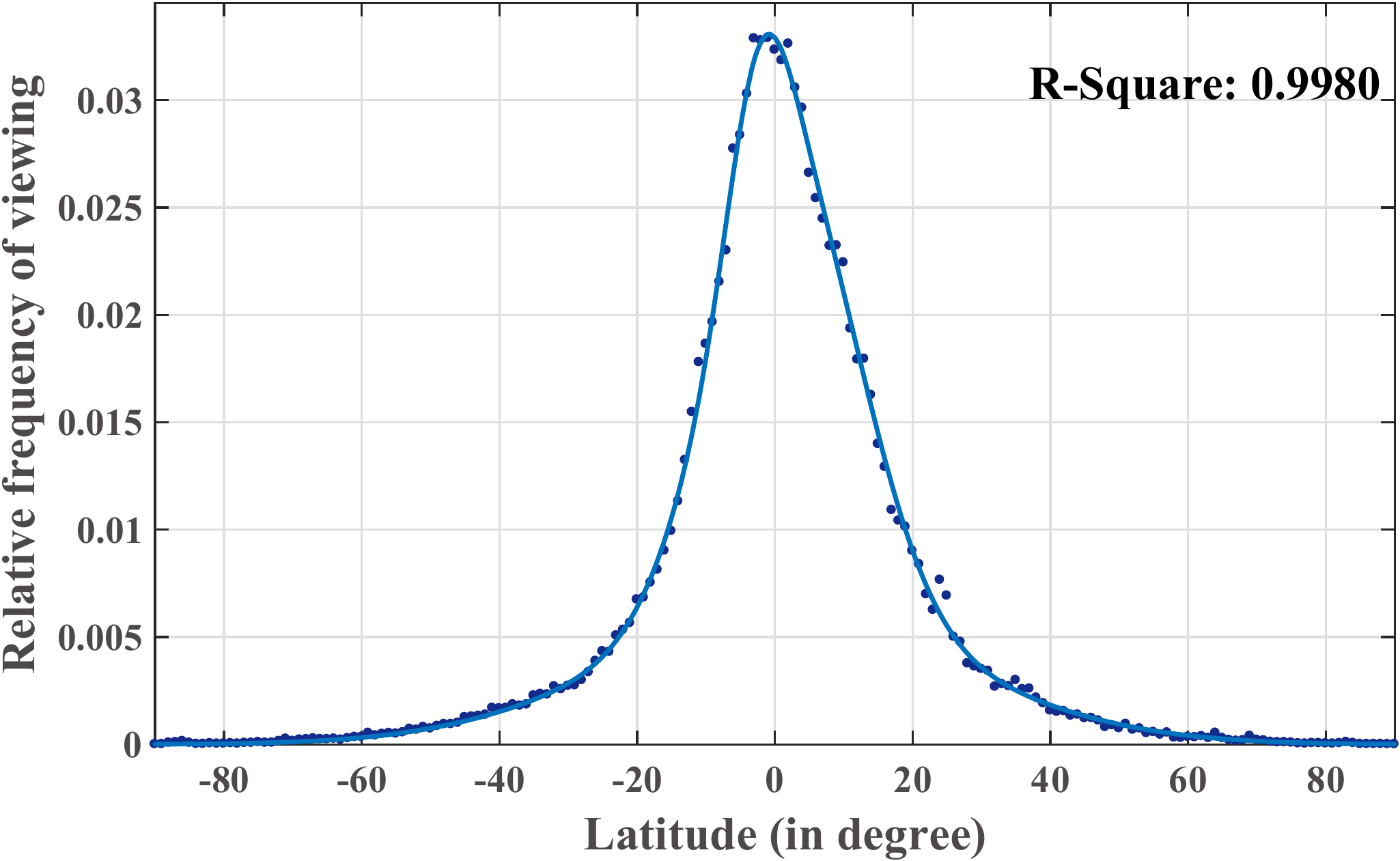}
}
\hspace{-1em}
\subfigure{
  \label{fig:frequency:b}
  \includegraphics[width=0.4\linewidth]{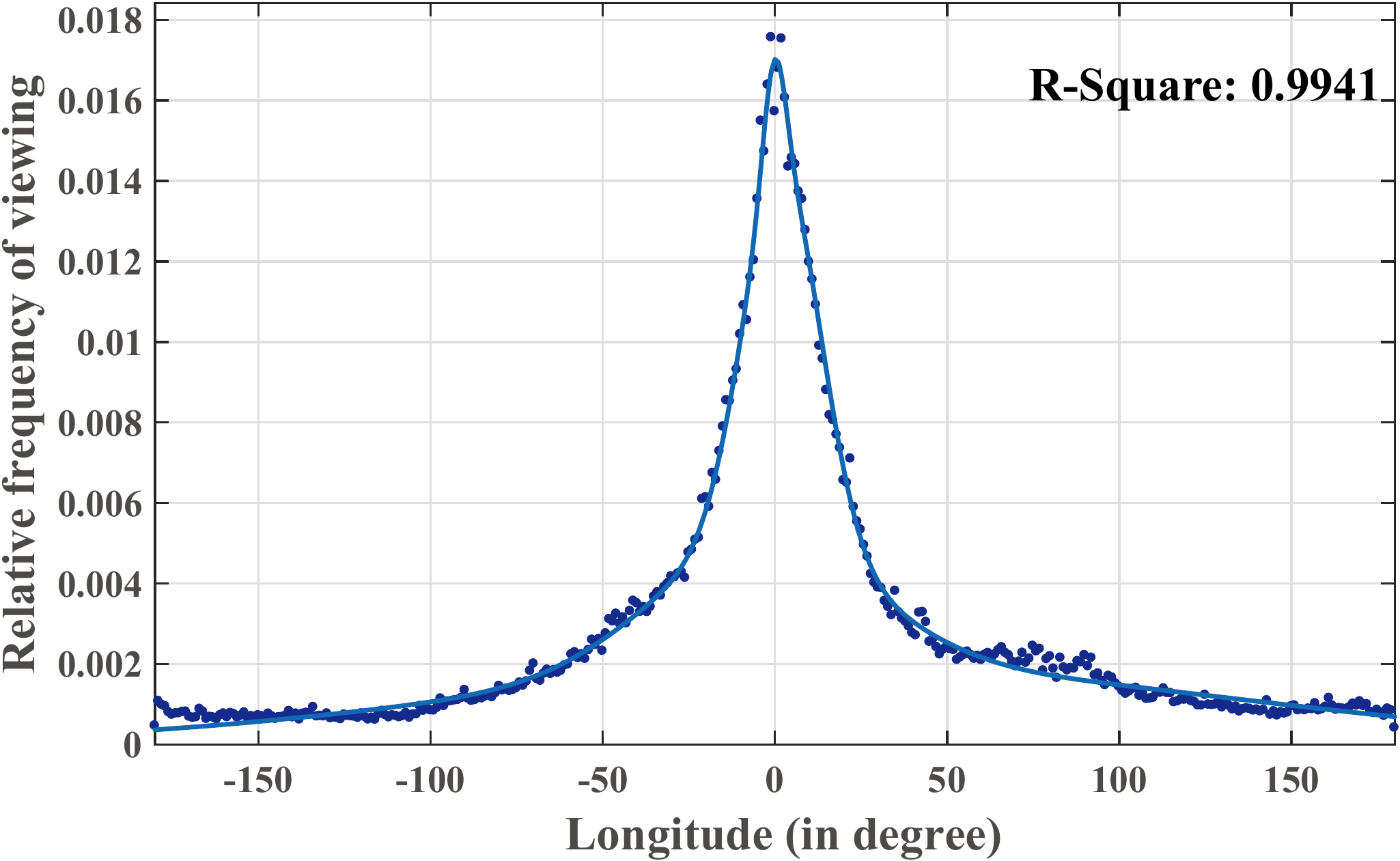}
}
\hspace{-0.07in}
}
\vspace{-2em}
\caption{Viewing direction frequency along the longitude and latitude.}
\label{fig:frequency}
\vspace{-0.8em}
\end{figure}

\textit{Finding 2: When watching omnidirectional video, subjects view the front region near the equator much more frequently than other regions.}

Figure \ref{fig:heatmap} shows the heat maps of viewing directions for some omnidirectional video sequences, as obtained from all 40 subjects. Note that the heat maps in Figure \ref{fig:heatmap} have been converted from spherical coordinates to a plane for omnidirectional video sequences \cite{snyder1987map}. We can see from this figure that most viewing directions fall into small regions located in the front region near the equator. Furthermore, we calculate the viewing directions belonging to different regions of the omnidirectional videos. Since \textit{Finding 1} illustrated that the longitude and latitude of viewing directions are almost uncorrelated with each other, it is reasonable to separately model the distribution of viewing directions along the longitude and latitude. To this end, Figure \ref{fig:frequency} shows the scatter diagrams of viewing direction frequency along the longitude and latitude, averaged over all subjects and all omnidirectional video sequences. In this figure, Gaussian mixture fitting curves are also plotted.
According to this figure, we can see that subjects tend to watch regions near the front and equator regions, far more often than the back and pole regions. This completes the analysis of \textit{Finding 2}, which is similar to the conclusion of \cite{yu2015framework}.
\begin{figure*}[!tb]
\begin{center}
\resizebox{\textwidth}{!}{
\hspace{-0.07in}
\subfigure[CandyCarnival]{
  \label{fig:inc-heatmap:a}
  \includegraphics[width=0.22\textwidth]{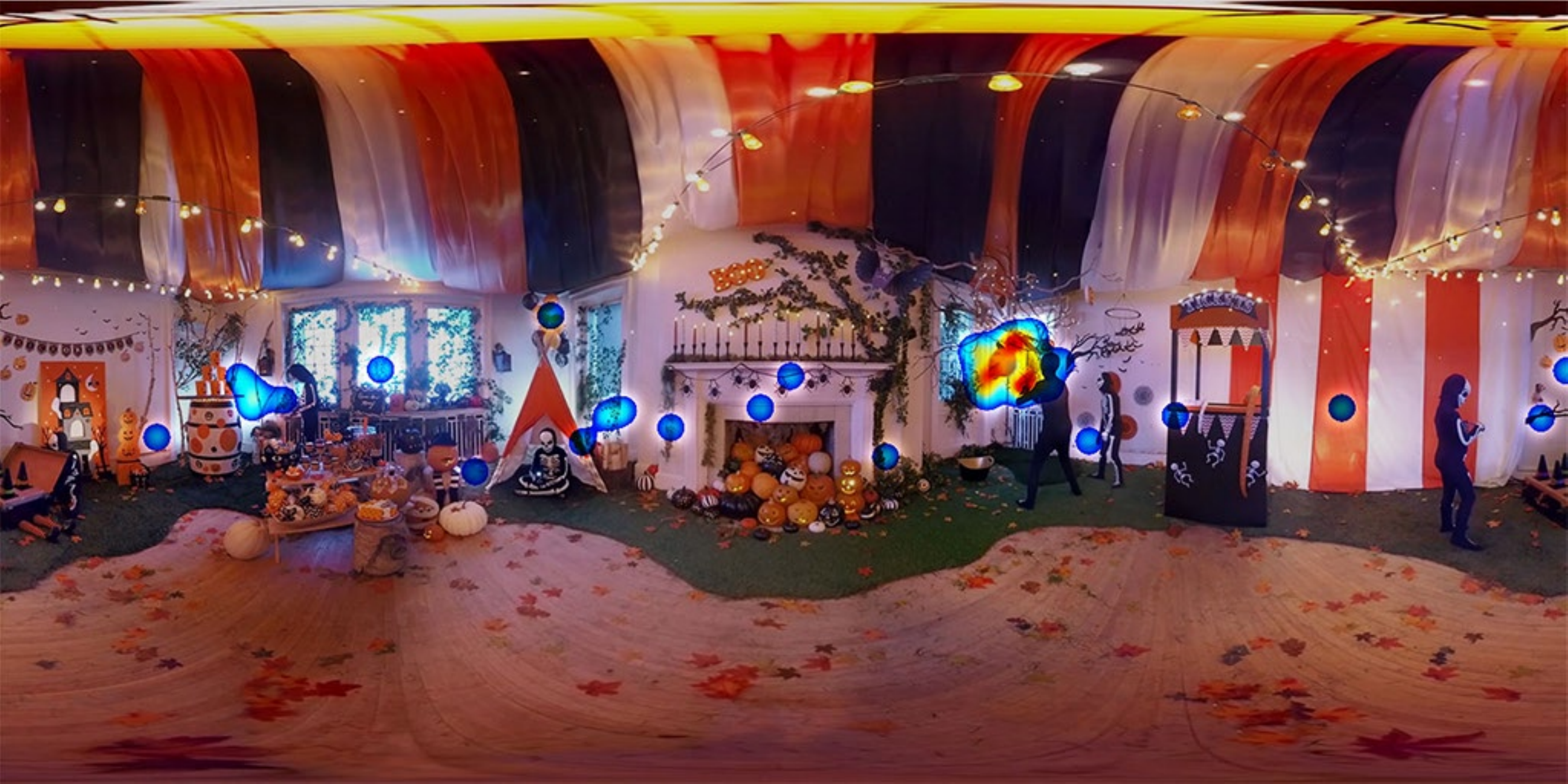}
}
\hspace{-0.1in}
\subfigure[MC]{
  \label{fig:inc-heatmap:b}
  \includegraphics[width=0.22\textwidth]{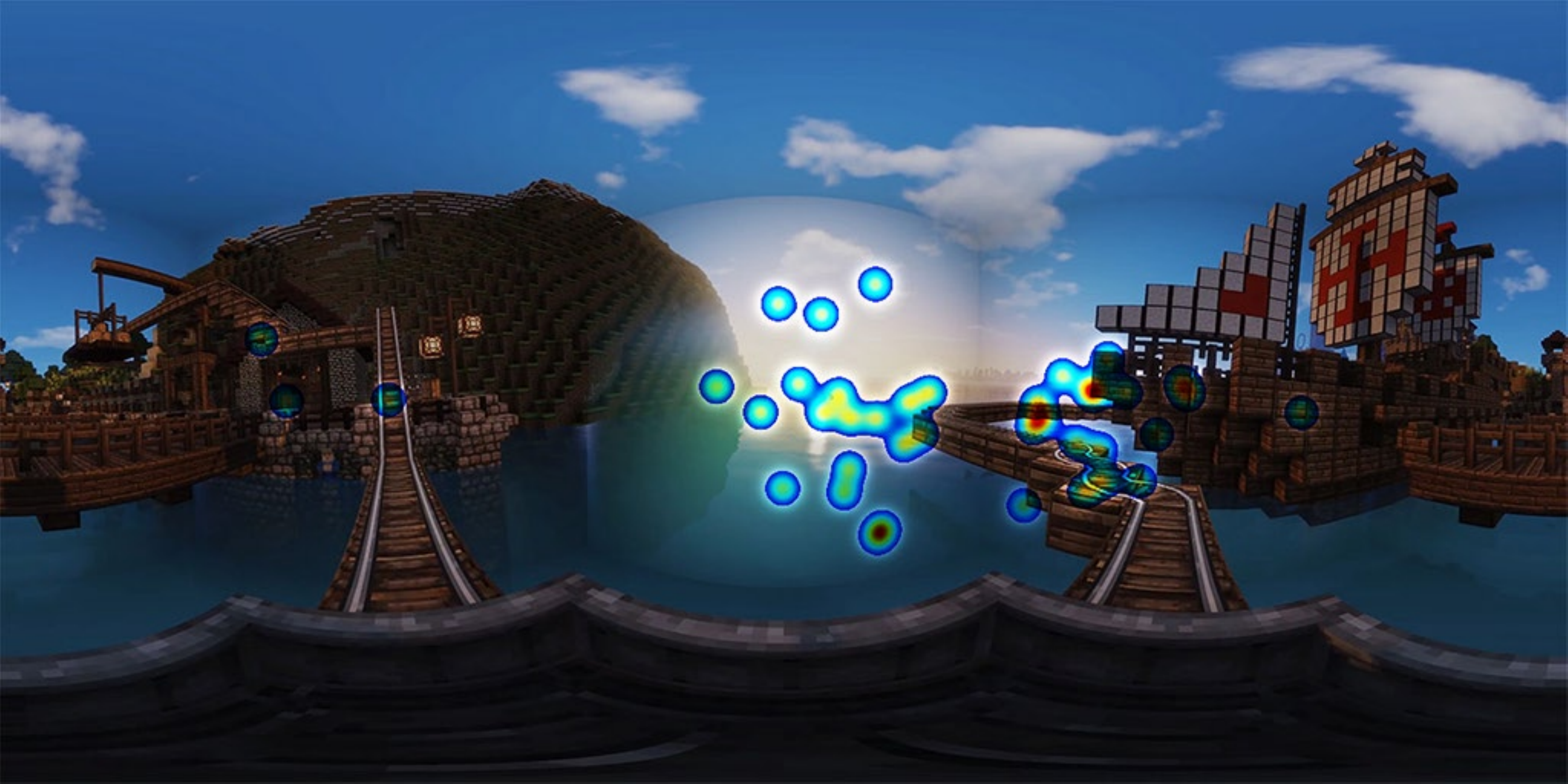}
}
\hspace{-0.1in}
\subfigure[Help]{
  \label{fig:inc-heatmap:c}
  \includegraphics[width=0.22\textwidth]{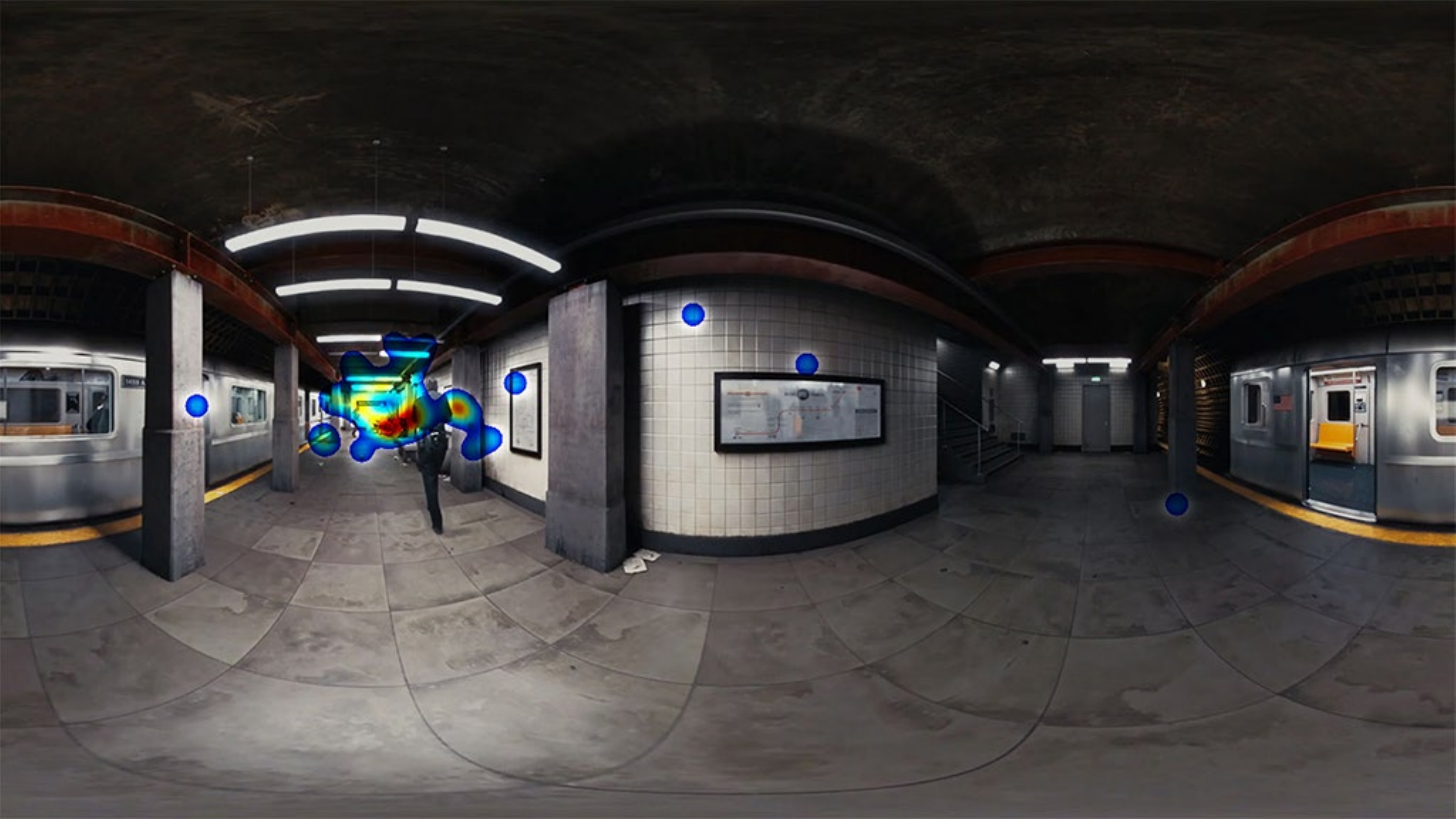}
}
\hspace{-0.1in}
\subfigure[StarryPolar]{
  \label{fig:inc-heatmap:d}
  \includegraphics[width=0.22\textwidth]{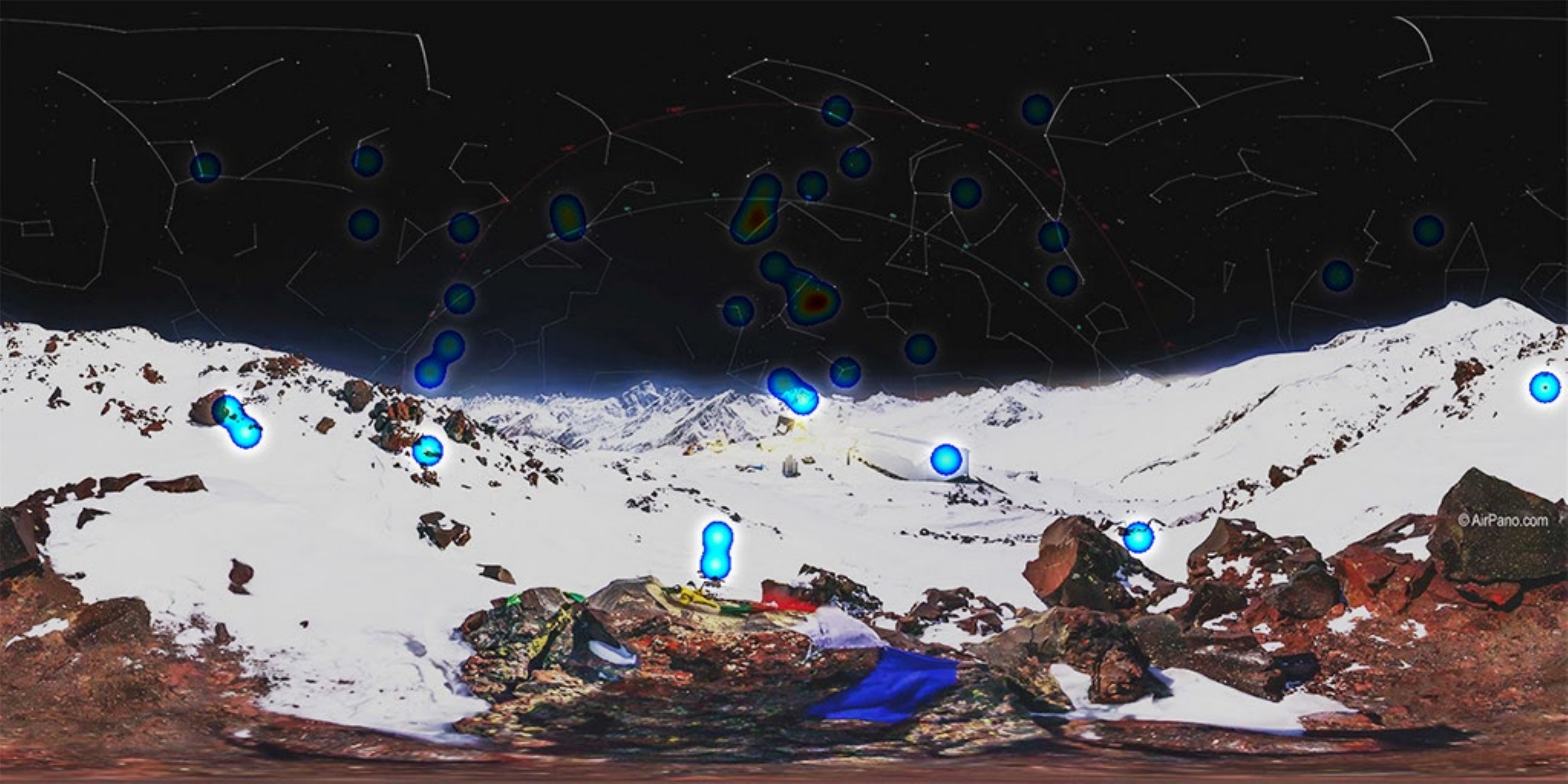}
}
\hspace{-0.1in}
\subfigure[NotBeAloneTonight]{
  \label{fig:inc-heatmap:e}
  \includegraphics[width=0.22\textwidth]{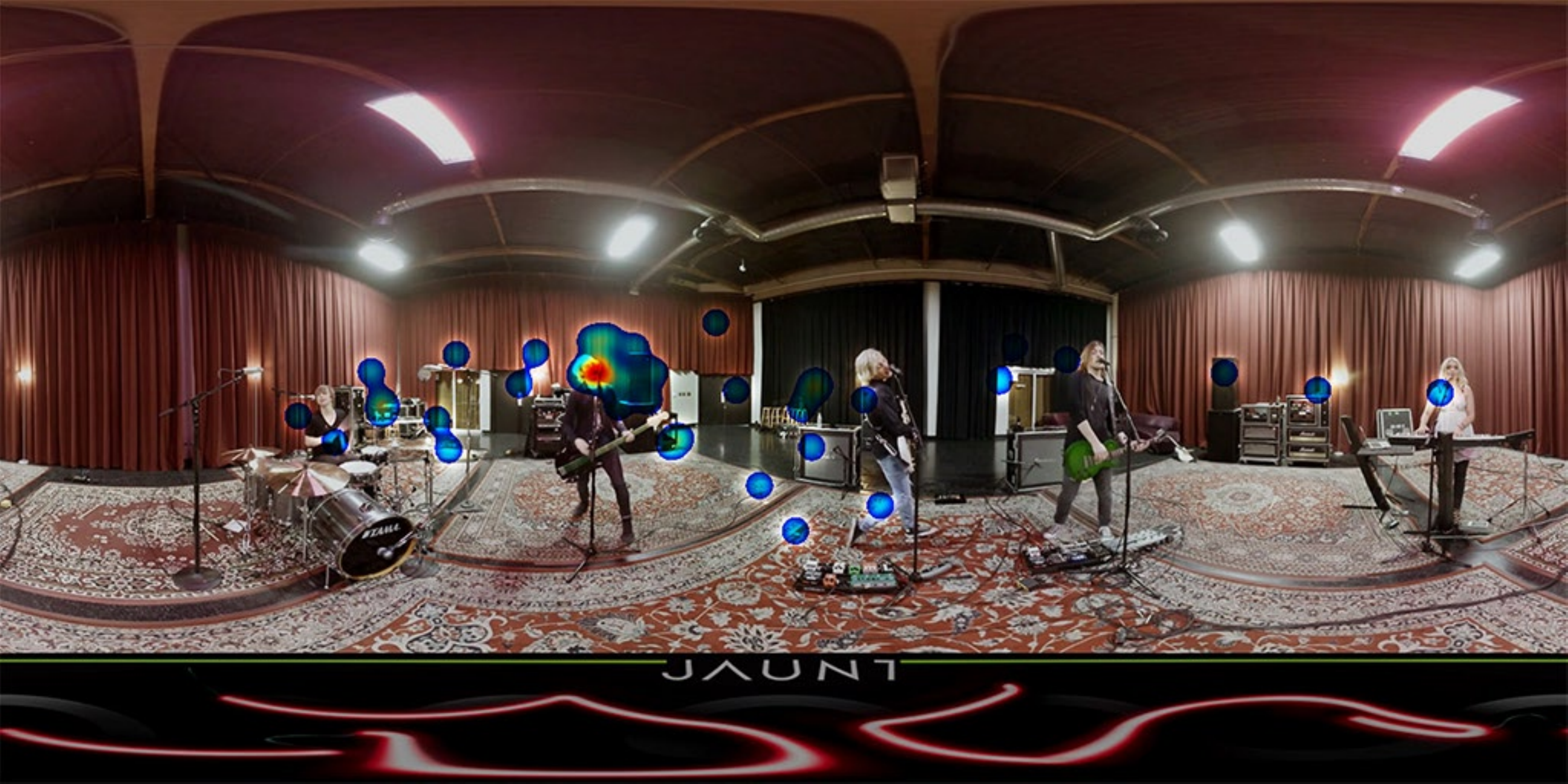}
}
\hspace{-0.07in}
}
\end{center}
\vspace{-1.3em}
\caption{Viewing direction heat maps for selected frames from a few omnidirectional video sequences, in which subjects are attracted by other regions.}
\label{fig:inc-heatmap}
\vspace{-0.8em}
\end{figure*}

\textit{Finding 3: In general, there exists high consistency in the viewed regions across different subjects for omnidirectional video.}

We randomly and equally divide all 40 subjects into two non-overlapping groups, $A$ and $B$. Then, we generate heat maps of viewing directions at one omnidirectional video frame for Groups $A$ and $B$, which are denoted as $\mathbf{H}_A$ and $\mathbf{H}_B$, respectively. Note that the heat maps for $\mathbf{H}_A$ and $\mathbf{H}_B$ are in plane coordinates, in which the omnidirectional video has been projected from sphere to plane. Here, we quantify the correlations of the heat maps of $\mathbf{H}_A$ and $\mathbf{H}_B$ using a linear correlation coefficient (CC) \cite{li2015data}. Specifically, CC is calculated by
\begin{equation}
\small
\begin{split}
&\mathrm{CC}(\mathbf{H}_A,\mathbf{H}_B) = \\ &\frac{\sum_{s,t}(\mathbf{H}_A(s,t)\!-\!\mu({\mathbf{H}_A}))\!\cdot\!(\mathbf{H}_B(s,t)\!-\!\mu({\mathbf{H}_B}))}{\sqrt{\sigma({\mathbf{H}_A})^2\cdot\sigma({\mathbf{H}_B})^2}}\mbox{,}
\end{split}
\end{equation}
where $(s,t)$ is the pixel coordinate, and $\mu(\cdot)$ and $\sigma(\cdot)$ are the mean and standard deviation of the corresponding heat maps, respectively. A CC (ranging in $[-1,1]$) close to $+1$ indicates a high consistency between the heat maps $\mathbf{H}_A$ and $\mathbf{H}_B$. Table~\ref{tab:CC} reports the mean values and standard deviations of the CC of the viewing direction heat maps for each sequence, between Groups $A$ and $B$. We can see from this table that the CC values are sufficiently high across different sequences. We can also see from this table that the CC value averaged over all 48 omnidirectional video sequences is $0.745$, with a standard deviation of 0.114. Thus, it is clear that the subjects behaved consistently when watching omnidirectional video. This completes the analysis of \textit{Finding 3}.

\textit{Finding 4: The viewing directions of different subjects are consistent in different regions according to content of omnidirectional video,
 despite being more likely to be attracted by equator and front regions.}

The scatter diagrams of Figure~\ref{fig:frequency} also reveal that the regions other than the front and equator, still have potential in attracting human attention. Figure~\ref{fig:inc-heatmap} demonstrates that the selected frames of several omnidirectional video sequences and their corresponding heat maps of viewing directions. We can see from Figure~\ref{fig:inc-heatmap} that the viewing directions may focus on different regions of omnidirectional video rather than the front equator, depending on the video content. For example, Figure~\ref{fig:inc-heatmap:c} shows that viewing directions concentrate on the corridor and people at the left hand side. This completes the analysis of this finding.
\section{Subjective VQA method}
\label{sec:method}
In this section, we introduce our subjective VQA method for omnidirectional video coding. In Section \ref{sec:general-arrangements}, we present the general configuration of the subjective test for our VQA method. In Section \ref{sec:test-procedure}, the procedure of the subjective test is discussed for rating the raw quality scores of each omnidirectional video sequence. In Section \ref{sec:processing}, O-DMOS and V-DMOS are proposed as the metrics to assess subjective quality of omnidirectional video coding, which are based on the raw scores of reference and impaired omnidirectional videos.
\subsection{Test configuration}\label{sec:general-arrangements}
Omnidirectional video differs from 2D video in the display devices, the viewing experience of subjects, etc. Thus, we design the test configuration for the subjective test on assessing omnidirectional video, which differs from the test for 2D video. In the following, we present the general configuration of the subjective test, including display devices and the setup for subjects.
\begin{figure}[!tb]
    \centering
    \resizebox{\linewidth}{!}{
    \subfigure[]{
    \label{fig:gui:original}
    \includegraphics[width=0.5\linewidth]{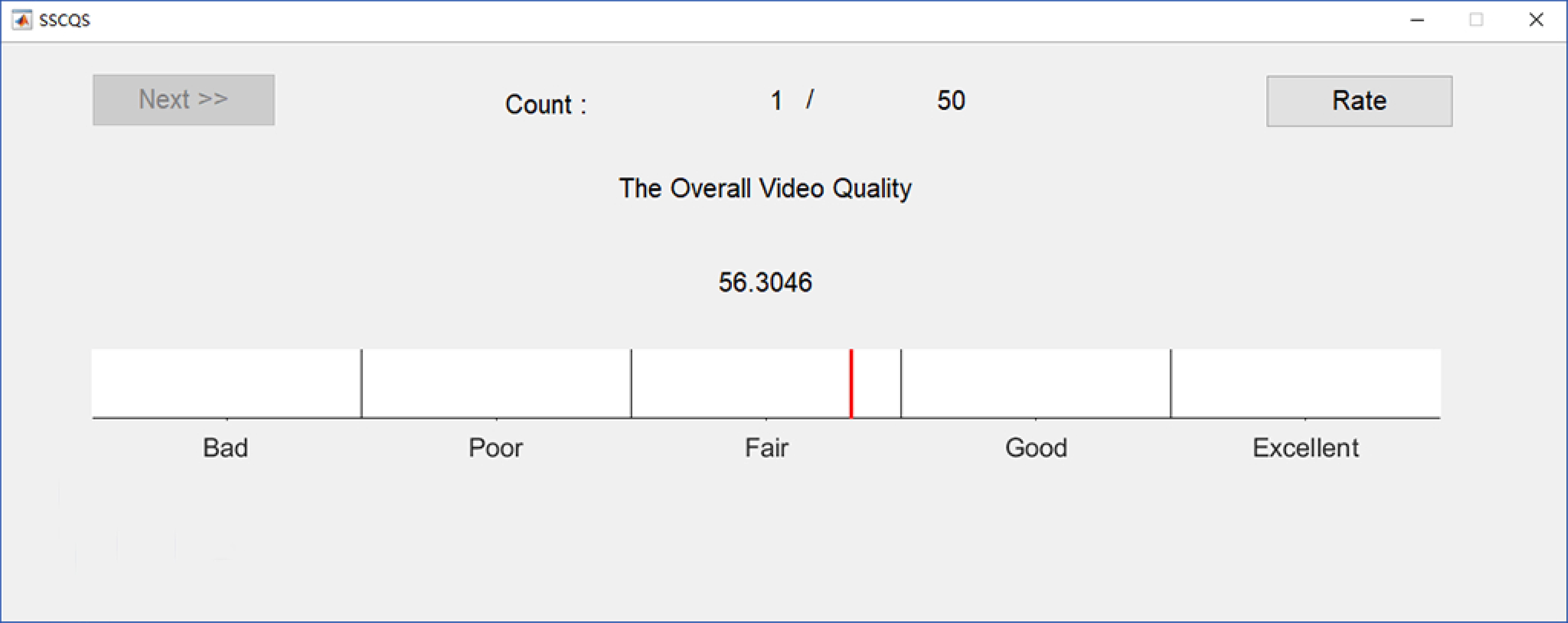}
    }
    \subfigure[]{
    \label{fig:gui:projected}
    \includegraphics[width=0.5\linewidth]{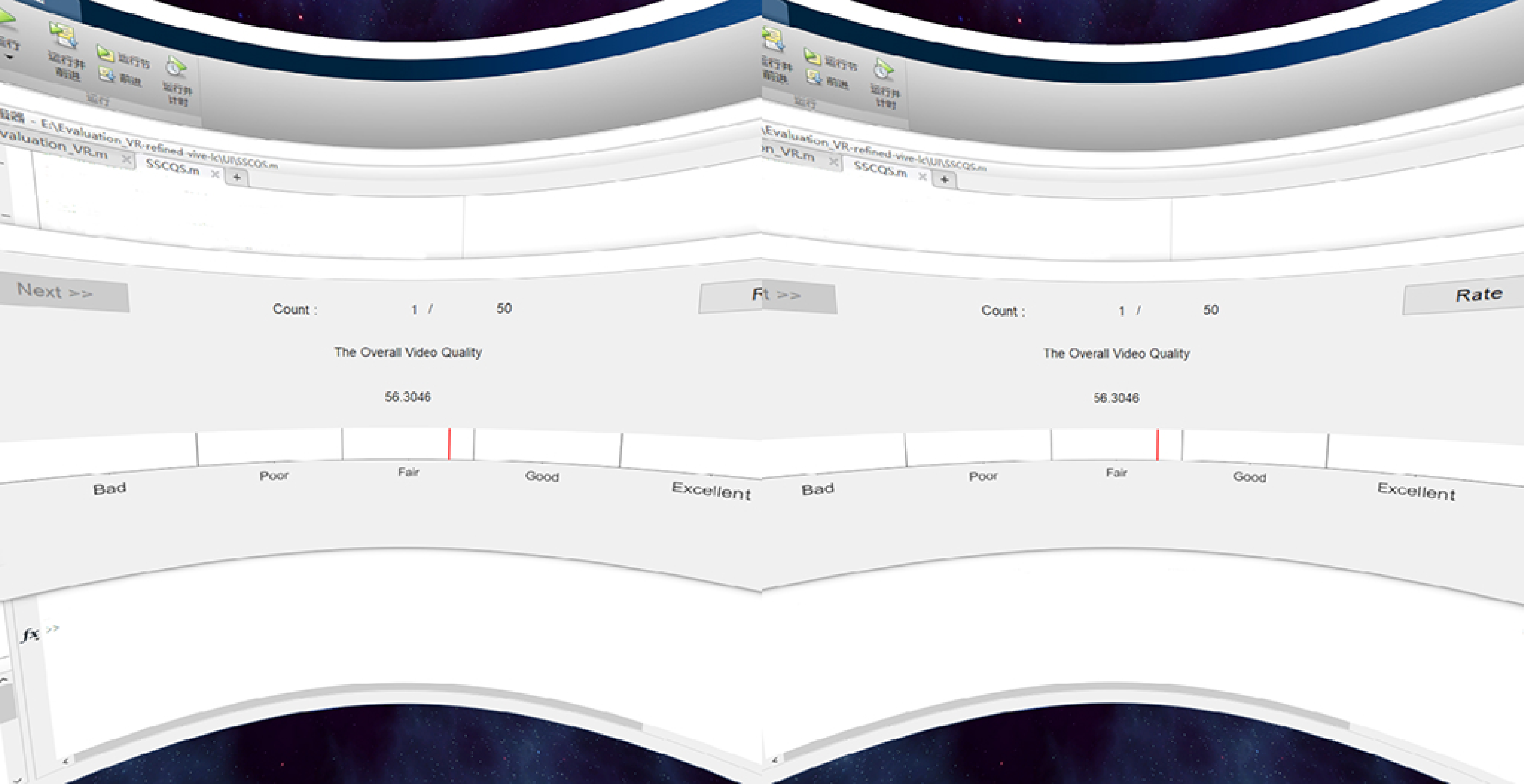}
    }
    }
    \vspace{-1.2em}
    \caption{(a) The GUI of our quality rating software. Subjects drag the red cursor on the continuous-scale slider with the controller to rate the scores. (b) The GUI in HMD projected by VD. Note that each of the left and right halves of the figure shows the picture in the corresponding eye in HMD, respectively.}
\label{fig:gui}
\vspace{-0.8em}
\end{figure}

\textbf{Display devices.} An HMD with a corresponding video player is used to display omnidirectional video, rather than flat screens for displaying 2D video. This configuration is because most omnidirectional videos are viewed by wearing an HMD. In this paper, we use the HTC Vive as the display device of HMD and the software VD as the omnidirectional video player.
Additionally, as shown in Figure \ref{fig:gui:projected}, VD is also used to project the graphical user interface (GUI) of our quality rating software in the HMD. This allows the subjects to rate omnidirectional video without taking off the HMD.
Since omnidirectional video can be viewed from different viewing directions, a swivel chair is provided to subjects when viewing the omnidirectional videos.

\textbf{Subjects.} According to \textit{Finding 3}, the viewing directions of subjects are highly consistent. Therefore, fixing the viewing regions of omnidirectional video in~\cite{zakharchenko2016quality} is not necessary. Instead, subjects are able to freely view all content of omnidirectional video in our subjective test. This way, our method satisfies daily visual experience that subjects are free to access all parts of omnidirectional video.
In addition, the initialization of viewing direction is required when watching omnidirectional video, which is different from viewing 2D video.
In our test, the viewing directions of all subjects are initialized to be the center of front region in omnidirectional video, as \textit{Finding 2} finds that subjects are more likely to be attracted by this region.
However, there still exists slight inconsistency of viewed regions in omnidirectional video as analyzed in \textit{Findings 4}. Thus, more subjects should be involved in the subjective test for rating quality of omnidirectional video, than at least 15 subjects required in~\cite{rec2012bt}. We recommend that at least 20 subjects are required for rating quality scores of omnidirectional video, as verified in Section~\ref{sec:perfomance}.
\subsection{Test procedure}
\label{sec:test-procedure}
\begin{figure}[!tb]
\centerline{\includegraphics[width=0.8\linewidth]{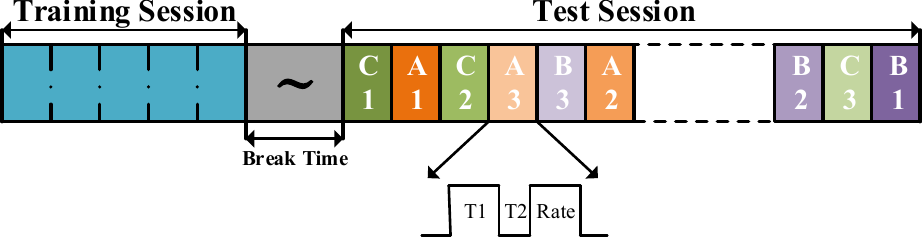}}
\vspace{-1em}
\caption{Structure of the test procedure with two sessions. The SSCQS procedure is illustrated in part of the test session. $Ai$, $Bi$ and $Ci$ represent various original and impaired sequences from different contents $A$, $B$ and $C$, respectively.}
\label{fig:sessions}
\vspace{-1em}
\end{figure}

\textbf{Training and test.} Generally speaking, the test procedure of our subjective VQA method comprises two sessions, the training and test sessions, as shown in Figure~\ref{fig:sessions}. The training session is introduced, as some subjects may be unfamiliar with viewing omnidirectional video. In the training session, subjects are told about the goal of our test. Then, they watch a group of training sequences at different quality levels in order to become familiar with omnidirectional video and its quality. Afterwards, a short break is required before entering the test session. In the test session, each sequence is displayed followed by a 3-second mid-grey screen. Compared with viewing 2D video, subjects are more likely to experience eye fatigue and motion sickness when watching omnidirectional video.
Thus, the maximum duration of a test session is limited to 30 minutes, which is the lower bound of recommended duration for 2D video in \cite{rec2012bt}.
If the test session lasts more than 30 minutes, a short break (at least 3 minutes) with the HMD taken off is added in the test session.

\textbf{Quality rating.} In the subjective test, SSCQS is adopted as shown in Figure~\ref{fig:sessions}, which means that omnidirectional video sequences are displayed in a random order and that sequences with the same content at different quality need to be avoided for two successive sequences. The reason for choosing SSCQS is that the subjects may continue to view unseen regions when viewing omnidirectional video with the same content, which differs from the viewing characteristics of 2D video.
After viewing each sequence, subjects are required to rate its quality. Note that there is one quality score to rate for each video.
As shown in Figure \ref{fig:gui:original}, the grading scores in the test session are available using a continuous-scale slider with a cursor in the GUI for quality rating.
The score $Q$ has a range from 0 to 100, in the form of 5 levels: excellent ($80\!\leq\!Q\!\leq\!100$), good ($60\!\leq\!Q\!<\!80$), fair ($40\!\leq\!Q\!<\!60$), poor ($20\!\leq\!Q\!<\!40$) and bad ($0\!\leq\!Q\!<\!20$).

\textbf{Data collection.} There are two kinds of data to be collected and processed, including the raw subjective quality scores of the omnidirectional video sequences as mentioned above. The other is the viewing direction data of subjects during sequence playback, which relate the quality score to the regions of omnidirectional video that were viewed. This also enables the calculation of V-DMOS, to be discussed next.
\subsection{Processing of subjective scores}
\label{sec:processing}
\textbf{O-DMOS.} Given the raw quality scores for each sequence, we follow the DMOS calculation method of 2D video as detailed in \cite{seshadrinathan2010study} to compute the O-DMOS, which indicates the overall quality of each omnidirectional video sequence. Specifically, the difference in the quality scores between the reference and impaired sequences is calculated for each subject. Let $S_{ij}$ and $S_{ij}^{\rm ref}$ denote the raw subjective scores assigned by subject $i$ to sequence $j$ and the corresponding reference sequence, respectively. Then, the difference score $d_{ij}$ can be simply obtained by
\begin{equation}
\small
d_{ij} = S_{ij}^{\mathrm{ref}}-S_{ij}\mbox{.}
\end{equation}
Afterwards, the difference score $d_{ij}$ needs to be converted to a Z-score $Z_{ij}$~\cite{van1995quality} using
\begin{equation}
\small
\mu_i=\frac{1}{M_i} \sum_{j=1}^{M_i} d_{ij}, \quad
\sigma_i=\sqrt{\frac{1}{M_i-1} \sum_{j=1}^{M_i} (d_{ij}-\mu_i)^2}\mbox{,}
\end{equation}
\begin{equation}
\small
\label{Zij}
Z_{ij}=\frac{d_{ij}-\mu_i}{\sigma_i}\mbox{,}
\end{equation}
where $M_i$ is the number of test sequences viewed by subject $i$.

Here, we need to ensure that each subject is valid by examining the Z-scores assigned by this subject. In other words, the Z-scores from invalid subjects should not be included when calculating the O-DMOS for measuring the subjective quality of omnidirectional video.
We apply the subject rejection method \cite{rec2012bt} to remove the Z-scores of some subjects if $5\%$ of the Z-scores assigned by these subjects fall outside the range of two standard deviations from the mean Z-scores.

Then, the Z-score $Z_{ij}$ needs to be linearly rescaled to fall within the range of $[0, 100]$. Assume that the Z-scores of a subject follow Gaussian distribution. Then, $Z_{ij}$ of \eqref{Zij} is distributed as a standard Gaussian, i.e. $\mathcal{N}(0,1)$, in which the mean is $0$ and the standard deviation is $1$. Thus, 99\% of the Z-scores lie in the range of $[-3, 3]$. To make such Z-scores $\in [0, 1]$, we normalize them by $(Z_{ij}+3)/6$. Then, the normalized Z-scores are rescaled to be $Z_{ij}^{'}$ as follows,
\begin{equation}
\small
Z_{ij}^{'}=\frac{100(Z_{ij}+3)}{6}\mbox{,}
\end{equation}
such that 99$\%$ of the values of $Z_{ij}^{'}$ fall into the range of $[0, 100]$.

Finally, the O-DMOS value of sequence $j$ is computed by averaging $Z_{ij}^{'}$ from $N_j$ valid subjects:
\begin{equation}
\small
{\mathrm{O\text{-}DMOS}}_j=\frac{1}{N_j} \sum_{i=1}^{N_j} Z_{ij}^{'}\mbox{.}
\end{equation}

\textbf{V-DMOS.} In addition to equirectangular projection (ERP), there are many other projections that map a sphere onto several planes of a polyhedron \cite{chen2018recent}, such as cube map projection (CMP) and octahedron projection. With these projection types, a whole omnidirectional video is split into parts, each of which is able to be compressed and transmitted separately. Therefore, in VQA of omnidirectional video, there is a need to split a whole omnidirectional video into several regions and then evaluate the quality of different regions. According to \textit{Finding 3}, there is still a slight inconsistency in the omnidirectional video viewing directions. \textit{Finding 4} further shows that all the regions of omnidirectional video can attract human attention. Thus, V-DMOS is used in our subjective VQA method to quantify the subjective quality of different regions of omnidirectional video, by making use of the collected raw quality scores and viewing direction data.

First, we need to compute the ratio of the frequency, with which subject $i$ views region $r$ in sequence $j$, denoted as $f_{ij}^r$. Note that $f_{ij}^r$ needs to be normalized to satisfy
\begin{equation}
\small
\sum_{r} f_{ij}^r =1 \mbox{.}
\end{equation}
When $f_{ij}^r>f_0$, where $f_0$ is a threshold, subject $i$ (after subject rejection~\cite{rec2012bt}) is added to collection $\mathbf{I}_{jr}$. Assuming that the size of $\mathbf{I}_{jr}$ is $N_{\mathbf{I}_{jr}}$, the DMOS value for region $r$ in sequence $j$ can be obtained by
\begin{equation}
\small
{\mathrm{DMOS}}_{jr}=\frac{1}{N_{\mathbf{I}_{jr}}} \sum_{i\in\mathbf{I}_{jr}} Z_{ij}^{'} \mbox{.}
\end{equation}
If $\mathbf{I}_{jr}=\emptyset$, then ${\rm DMOS}_{jr}$ is an invalid value, denoted by ``---''.
Finally, the vector of V-DMOS can be represented by
\begin{eqnarray}
\small
\begin{bmatrix}
{\mathrm{O\text{-}DMOS}}_j & {\mathrm{DMOS}}_{j1} & \cdots & {\mathrm{DMOS}}_{jr} & \cdots & {\mathrm{DMOS}}_{jR}
\end{bmatrix}
\mbox{,}
\end{eqnarray}
where $R$ is the total number of regions in omnidirectional video.
The sphere is split into 6 regions in the same way as the CMP \cite{dimitrijevic2016comparison}, in which the front, left, right, back, top and bottom regions are extracted according to the longitude and latitude. For more details, refer to \cite{dimitrijevic2016comparison}.

As a result, our V-DMOS is able to measure both overall and regional quality degradation for impaired omnidirectional video. There exist some works on region/viewport oriented coding optimization \cite{tang2017optimized,li2017projection,he2017motion} and region/viewport adaptive streaming \cite{sreedhar2016viewport,taghavinasrabadi2017adaptive,budagavi2015360}. The optimization or adaptation schemes consider inequality of different regions. With the score vector of V-DMOS, we can directly evaluate the performance of the schemes designed for specific regions of omnidirectional video.
\section{Objective VQA methods}
In this section, we propose two objective VQA methods for omnidirectional video coding, which are on the basis of the traditional PSNR mechanism and our findings in Section~\ref{sec:data-analysis}. Both of these methods impose weights on the pixel-wise distortion in calculating the PSNR, according to the possibility of attracting human attention. Thus, these methods are called perceptual VQA (P-VQA) methods.
The first method mainly focuses on weighting the distortion of pixels according to their locations in omnidirectional video rather than their contents. Thus, this method is called the non-content-based P-VQA (NCP-VQA) method, to be discussed in Section \ref{sec:non-content}. The second method allocates weights to pixel-wise distortion based on the viewing directions predicted with respect to the content of omnidirectional video and is thus called the content-based P-VQA (CP-VQA) method. This is to be introduced in Section \ref{sec:content}.
\begin{table}
\caption{Values of the parameters in~\eqref{eq:map}.} \label{tab:coef-value}
\vspace{-2em}
\begin{center}
\begin{tabular}{*{7}{|c}|}
  \hline
  $k$/$k'$ & $a_k$ & $b_k$ & $c_k$ & $a'_{k'}$ & $b'_{k'}$ & $c'_{k'}$ \\
  \hline
  1 & 0.0034 & -0.1549 & 4.6740 & 0.0075 & -2.3738 & 6.6437 \\
  \hline
  2 & 0.0106 & 1.5140 & 18.51 & 0.0209 & 1.8260 & 14.8171 \\
  \hline
  3 & 0.0032 & 6.3670 & 110.5 & 0.0057 & 1.4618 & 36.1311 \\
  \hline
\end{tabular}
\end{center}
\vspace{-2.5em}
\end{table}
\subsection{Non-content-based perceptual VQA method}\label{sec:non-content}
According to \textit{Finding 2}, front regions near the equator are viewed more frequently than other regions in omnidirectional video. Thus, it is necessary to consider such viewing direction frequency when calculating the non-content-based perceptual PSNR (NCP-PSNR) for our NCP-VQA method for omnidirectional video coding. Let $\varphi \in[-180^{\circ},180^{\circ}]$ and $\theta \in[-90^{\circ},90^{\circ}]$ denote the longitude and latitude of a viewing direction in degrees, respectively. Since \textit{Finding 1} points out that the longitude and latitude of the viewing directions are almost independent of each other, the distribution of viewing direction frequency $u(\varphi,\theta)$ can be modeled using the following Gaussian mixture model (GMM):
\begin{equation}
\small
\label{eq:map}
\begin{split}
&u(\varphi,\theta)= \\
&\!\underbrace{\left\{\!\sum_{k=1}^{3}{\!a_k\exp\!\left[\!-\!\left(\!\frac{\varphi\!-\!b_k}{c_k}\!\right)^2\!\right]}\!\right\}}_{\text{GMM for longitude}}\! \underbrace{\left\{\!\sum_{k'=1}^{3}{\!a'_{k'}\exp\!\left[\!-\!\left(\!\frac{\theta\!-\!b'_{k'}}{c'_{k'}}\!\right)^2\!\right]}\!\right\}}_{\text{GMM for latitude}}\mbox{.}
\end{split}
\end{equation}
\begin{figure*}[!tb]
\begin{center}
\subfigure[Block diagram of the content-based PVQA method.]{
  \label{fig:cpvqa:a}
  \includegraphics[width=0.8\textwidth]{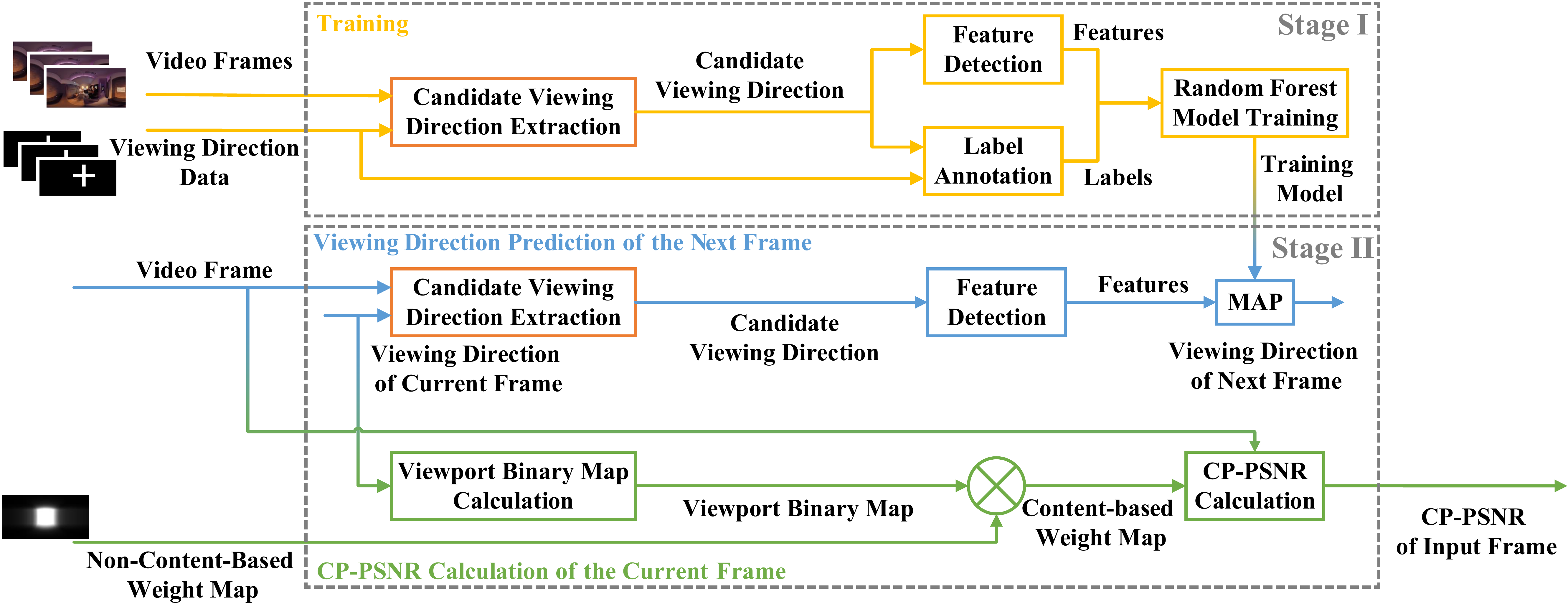}
}
\subfigure[Detail of candidate viewing direction extractor.]{
  \label{fig:cpvqa:b}
  \includegraphics[width=0.8\textwidth]{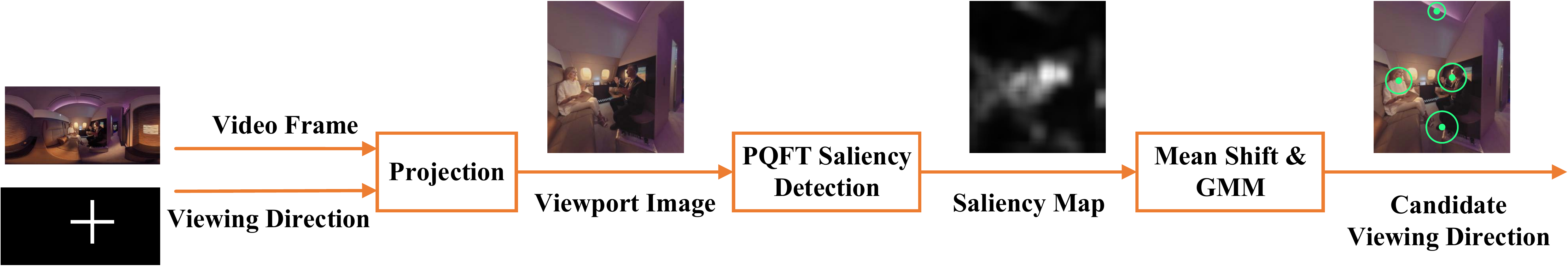}
}
\end{center}
\vspace{-1.5em}
\caption{The procedure of the proposed content-based PVQA method.}
\vspace{-0.8em}
\label{fig:cpvqa}
\end{figure*}
%
%
In \eqref{eq:map}, $a_k$, $b_k$ and $c_k$ are parameters of the GMM for the viewing direction distribution of longitude, and $a'_{k'}$, $b'_{k'}$ and $c'_{k'}$ are GMM parameters for the viewing direction distribution of latitude.
The values of these parameters can be obtained via least squares fitting over all viewing directions of our database proposed in Section \ref{sec:database}, and they are reported in Table \ref{tab:coef-value}\footnote{The GMM model with these parameters is verified highly correlated with the ground truth viewing direction data of the test set of in Section \ref{sec:experiment}.}.
In fact, this least squares fitting has already been discussed in the analysis of \textit{Finding 2}, with GMM curves been plotted in Figure 2.
Note that the numbers of Gaussian components for both GMM fitting of \eqref{eq:map} are set to 3, for making the fitting error convergent (with R-square values more than 0.99).

Next, we take into account the equirectangular projection for our NCP-PSNR metric. Given an omnidirectional video under an equirectangular projection with a resolution of $W\times H$, the probability of each pixel being in the viewing direction during one frame can be obtained by \cite{snyder1987map}:
\begin{equation}
\small
\label{eq:fxy}
\begin{split}
&v(s,t)= \\
&u\!\left(\!-360\left(\frac{s-1}{W-1}\!-\!\frac{1}{2}\right)\!,\!-180\left(\frac{t-1}{H-1}\!-\!\frac{1}{2}\right)\!\!\right)\!\mbox{,}
\end{split}
\end{equation}
where $(s,t)$ is the pixel coordinate, with $1\le s \le W$ and $1\le t \le H$. Note that our VQA method can be easily extended to other projections by replacing the projection formulation.

In fact, pixels within a viewport centered in one viewing direction are all accessible to subjects. Thus, the pixels within a viewport should be of equal importance in evaluating the quality of encoded omnidirectional video. To model possible viewports, we need to generate the non-content-based weight map in our NCP-VQA method, based on the probability of viewing direction, i.e., $v(s,t)$ in \eqref{eq:fxy}. Assume that $\mathbf{P}_{s,t}$ is the viewport, the center of which is the viewing direction $(s,t)$. According to studies on near peripheral vision \cite{besharse2011retina}, the ranges of $\mathbf{P}_{s,t}$ are set to $[-30^{\circ}, 30^{\circ}]$ in both directions. For each pixel $(s,t)$ in an omnidirectional video frame, we can find all viewports including this pixel, and the viewing directions of these viewports constitute a collection denoted by $\mathbf{V}_{s,t}$. Then, the non-content-based weight map can be obtained by
\begin{equation}
\small
w(s,t)=\mathop{\max}_{(s',t')\in\mathbf{V}_{s,t}}v(s',t')\mbox{.}
\end{equation}

Afterwards, the non-content-based weight map needs to be normalized by
\begin{equation}
\small
\label{eq:wtilde}
\tilde{w}(s,t)=\frac{w(s,t)}{\sum_{s,t}{w(s,t)}},
\end{equation}
to satisfy
\begin{equation}
\small
\label{eq:wsum}
\sum_{s,t}\tilde{w}(s,t)=1\mbox{.}
\end{equation}
Note that the non-content-based weight map can be trained offline. Then, the sequences with the same resolution can use the same offline weight map, without any extra complexity. Finally, based on the definition of the PSNR, the NCP-PSNR for each omnidirectional video frame can be calculated\footnote{Because of \eqref{eq:wtilde} and \eqref{eq:wsum}, we do not need to divide NCP-PSNR in \eqref{eq:NC-VQA} by the number of pixels.} as
\begin{equation}
\label{eq:NC-VQA}
\small
\text{NCP-PSNR} = 10\log\frac{I_{\mathrm{max}}^2}{\sum_{s,t}{(I(s,t)\!-\!I'(s,t))^2\!\cdot\!\tilde{w}(s,t)}}\mbox{,}
\end{equation}
where $I(s,t)$ and $I'(s,t)$ are intensities of pixel $(s,t)$ in the reference and processed omnidirectional videos, respectively. Additionally, $I_{\mathrm{max}}$ is the maximum intensity value of the videos (= 255 for 8-bit intensity).

\subsection{Content-based PVQA method}\label{sec:content}
\textit{Finding 4} shows that the viewing directions of the subjects are also correlated with the contents of the omnidirectional video. According to this finding, we further develop a CP-VQA method for omnidirectional video coding, in which the content-based perceptual PSNR (CP-PSNR) is measured.

\textbf{Outline}. Figure~\ref{fig:cpvqa:a} summarizes the procedure of our CP-VQA method. As shown in this figure, our CP-VQA method consists of two stages, i.e., Stage I: model training, and Stage II: CP-PSNR calculation.
For Stage I of model training, the input includes omnidirectional video frames and their corresponding viewing directions for all subjects.
Then, a random forest model of classification is trained to predict viewing directions.
For Stage II of CP-PSNR calculation, each omnidirectional video frame is taken as the input.
In this stage, there are two branches to accomplish two different tasks. (1) Viewing direction prediction of the next frame. After extracting several candidates from the input frame, a viewing direction can be predicted using maximum a posteriori (MAP) estimation for the incoming frames, with regard to detected features and the trained model.
(2) CP-PSNR calculation of the current frame. Given the predicted viewing direction, the viewport binary map is calculated. Then, the content-based weight map is generated by multiplying the viewport binary map with the non-content-based weight map. Finally, the CP-PSNR is obtained by imposing the content-based weight map in the PSNR. In the following, we present the details of our CP-VQA method.

\textbf{Extraction of candidate viewing directions}. In our CP-VQA method, the first step is to extract viewing direction candidates for the next omnidirectional video frame.
Given the input omnidirectional video frame and its current viewing direction, the procedure for extracting the viewing direction candidates is shown in Figure \ref{fig:cpvqa:b}. This procedure also corresponds to the orange blocks ``candidate viewing direction extraction'' in Figure \ref{fig:cpvqa:a}.
To extract the viewing direction candidates, a viewport projection \cite{yu2015framework} is applied to obtain the viewport image of the input video frame given the current viewing direction. Note that the viewport image is regarded as the image that subjects can actually see through the HMD with little affine geometric distortion. Therefore, the feature extraction and saliency prediction methods for 2D image are applicable to the viewport image. Subsequently, a saliency map of the viewport image is generated using a simple yet effective saliency detection algorithm, phase spectrum of quaternion Fourier transform (PQFT) \cite{guo2010novel}.
Considering the saliency map as a 2-dimensional probability distribution, ten thousand of random points subject to this distribution can be generated. Then, the mean-shift \cite{cheng1995mean} and GMM are applied on these random points for clustering. The center of each cluster is used as a candidate of the viewing direction for the incoming frame (the green dots in Figure \ref{fig:cpvqa:b}). The standard deviation parameter of each GMM represents the area of the salient region around the candidate (the green circles around the dots in Figure \ref{fig:cpvqa:b}).

\textbf{Stage I: Model training}. Regarding the model training, the inputs are video frames and their corresponding viewing directions for each subject, which are derived from our viewing direction database for omnidirectional video. Then, several candidates of viewing directions are obtained using the aforementioned extractor. For each candidate, the features, which are correlated with the viewing direction transition from the current one to the candidate, are detected to train a random forest classifier~\cite{liaw2002classification}. Here, we select the features of 2D image saliency prediction, since \cite{rudoy2013learning} validates that these features are effective in predicting human attention. The features include (1) the Euclidean distance from the viewport center to the candidate; (2) the angle between the viewport center and the candidate; (3) the standard deviation of the GMM used for extracting the candidate; (4) the averaged saliency value of the region around the candidate; and (5) the local intensity contrast of the neighborhood around the candidate \cite{rudoy2013learning}. These features form a vector $\bm{\upsilon}$, which is the input to the random forest classifier. Furthermore, the viewing direction for the same subject in the next frame is annotated as the ground-truth and is the target output of the random forest classifier. Finally, the random forest classifier can be trained with the feature vectors and ground-truth viewing directions, and is then used to calculate the CP-PSNR of each omnidirectional video frame. Note that this stage is run offline only once to obtain the trained random forest model.
%

\textbf{Branch 1, Stage II: Viewing direction prediction of the next fame}. For the CP-PSNR calculation, the first step is to predict the viewing direction. To predict the viewing direction, a few viewing direction candidates are extracted. Then, the MAP estimation is employed to select one viewing direction from the candidates given the detected features embedded in vector $\bm{\upsilon}$ and the trained random forest model. Specifically, assuming that $C$ is a viewing direction candidate, the averaged posterior probability of candidate $C$ being the viewing direction (i.e., belonging to the positive class) can be obtained by
{
\begin{equation}
\small
g_{\lambda_{+}}(C)=\frac{1}{T} \sum_{\tau=1}^{T} P(\lambda_{+}|\bm{\upsilon}_{\tau}(C))\mbox{,}
\end{equation}
where $\lambda_{+}$ represents the positive class, and $\bm{\upsilon}_{\tau}(C)$ is the feature vector of $C$ input to tree $\tau$. In addition, $T$ is the number of classification trees in the trained random forest model. Note that each tree randomly chooses some features from the input feature vector $\bm{\upsilon}$ for the classification, such that $\bm{\upsilon}_{\tau}(C) \subseteq \bm{\upsilon}$. Finally, the viewing direction is predicted for the next frame by MAP as follows,
{\setlength\abovedisplayskip{1pt}
\setlength\belowdisplayskip{1pt}
\begin{equation}
\small
V = \argmax_{C} g_{\lambda_{+}}(C)\mbox{.}
\end{equation}}
%

\textbf{Branch 2, Stage II: CP-PSNR calculation of the current frame}.
Given the predicted viewing direction of the current frame, a viewport binary map can be generated, in which 1 indicates that the corresponding pixel is in the viewport range and 0 means that the pixel is out of the viewport range.
Then, a content-based weight map $w'(s,t)$ is generated via multiplying the viewport binary map by the non-content-base weight map $\tilde{w}(s,t)$ (introduced in Section \ref{sec:non-content}). We further normalize $w'(s,t)$ by
\begin{equation}
\label{eq:wtilde2}
\small
\widetilde{w'}(s,t)=\frac{w'(s,t)}{\sum_{s,t}{w'(s,t)}}\mbox{.}
\end{equation}
Finally, the CP-PSNR of each omnidirectional video frame can be calculated as
\begin{equation}
\label{eq:C-VQA}
\small
\text{CP-PSNR} = 10\log\frac{I_{\mathrm{max}}^2}{\sum_{s,t}{(I(s,t)-I'(s,t))^2 \cdot \widetilde{w'}(s,t)}}\mbox{.}
\end{equation}
As a result, CP-VQA can be obtained for measuring the objective quality of omnidirectional video.

%
\begin{table*}[!tb]
\begin{center}
  \caption{The final output as the V-DMOS of the impaired test sequences.}
  \label{tab:dmos}
  \vspace{-2em}
  \resizebox{\textwidth}{!}{
  \begin{tabular}{*{9}{|c}|}
  \hline
  QP & Name & V-DMOS & Name & V-DMOS & Name & V-DMOS & Name & V-DMOS \\
  \hline
  27 &\multirow{3}{*}{Dianying} & [\textbf{43},43,45,36,43,48,33] &
  \multirow{3}{*}{Fengjing1} & [\textbf{36},37,37,36,34,---,---] &
  \multirow{3}{*}{Fengjing3} & [\textbf{36},35,34,43,38,72,21] &
  \multirow{3}{*}{Hangpai1} & [\textbf{33},33,30,28,29,---,34] \\
  \cline{1-1} \cline{3-3} \cline{5-5} \cline{7-7} \cline{9-9}
  37 & & [\textbf{65},65,64,69,66,---,57] & & [\textbf{64},64,65,70,66,64,---] & & [\textbf{43},44,47,41,47,---,35] & & [\textbf{47},47,48,41,46,---,41] \\
  \cline{1-1} \cline{3-3} \cline{5-5} \cline{7-7} \cline{9-9}
  42 & & [\textbf{71},71,66,64,71,---,54] & & [\textbf{70},70,70,60,70,---,55] & & [\textbf{54},55,54,44,54,---,38] & & [\textbf{58},58,53,65,52,---,51] \\
  \hline
  27 &\multirow{3}{*}{Hangpai2} & [\textbf{33},33,32,34,34,---,19] &
  \multirow{3}{*}{Hangpai3} & [\textbf{36},36,35,33,36,---,---] &
  \multirow{3}{*}{Tiyu1} & [\textbf{43},43,40,42,39,---,---] &
  \multirow{3}{*}{Tiyu2} & [\textbf{35},35,31,---,30,---,---] \\
  \cline{1-1} \cline{3-3} \cline{5-5} \cline{7-7} \cline{9-9}
  37 & & [\textbf{40},40,40,43,40,---,32] & & [\textbf{47},47,44,48,48,---,46] & & [\textbf{59},59,56,---,60,---,66] & & [\textbf{58},57,59,60,61,---,70] \\
  \cline{1-1} \cline{3-3} \cline{5-5} \cline{7-7} \cline{9-9}
  42 & & [\textbf{52},52,54,37,51,---,48] & & [\textbf{58},57,55,63,62,---,---] & & [\textbf{70},71,66,67,65,55,---] & & [\textbf{66},66,64,---,66,---,71] \\
  \hline
  27&\multirow{3}{*}{Xinwen1} & [\textbf{34},33,33,34,33,---,35] &
  \multirow{3}{*}{Xinwen2} & [\textbf{34},34,33,34,34,---,46] &
  \multirow{3}{*}{Yanchanghui1} & [\textbf{35},34,35,42,34,---,---] &
  \multirow{3}{*}{Yanchanghui2} & [\textbf{34},33,33,32,34,---,---] \\
  \cline{1-1} \cline{3-3} \cline{5-5} \cline{7-7} \cline{9-9}
  37 & & [\textbf{47},46,48,47,47,---,---] & & [\textbf{55},55,55,51,56,---,59] & & [\textbf{45},43,46,59,43,43,---] & & [\textbf{52},50,52,58,55,---,---] \\
  \cline{1-1} \cline{3-3} \cline{5-5} \cline{7-7} \cline{9-9}
  42 & & [\textbf{58},57,61,72,57,---,50] & & [\textbf{67},67,66,59,67,---,70] & & [\textbf{59},58,62,65,58,---,---] & & [\textbf{62},62,62,58,64,63,---] \\
  \hline
  \end{tabular}}
\end{center}
\vspace{-2em}
\end{table*}
\begin{figure}[!tb]
    \centering
    \subfigure[Reference]{
    \label{fig:qp:0}
    \includegraphics[width=0.45\linewidth]{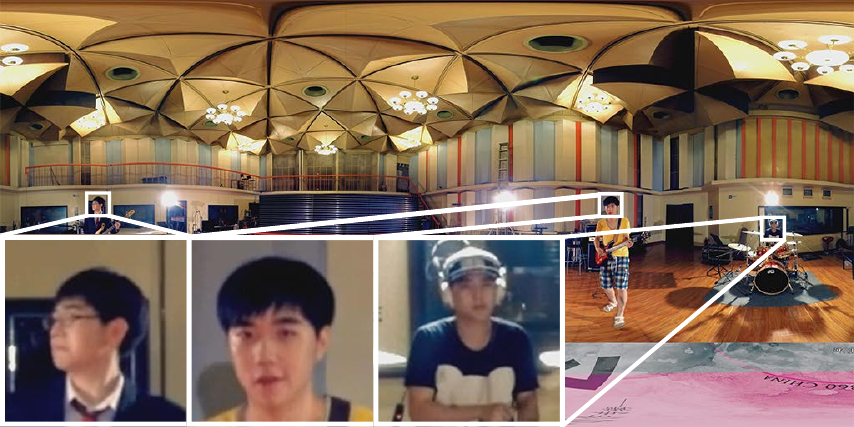}
    }\hspace{-0.5em}
    \subfigure[QP=27]{
    \label{fig:qp:27}
    \includegraphics[width=0.45\linewidth]{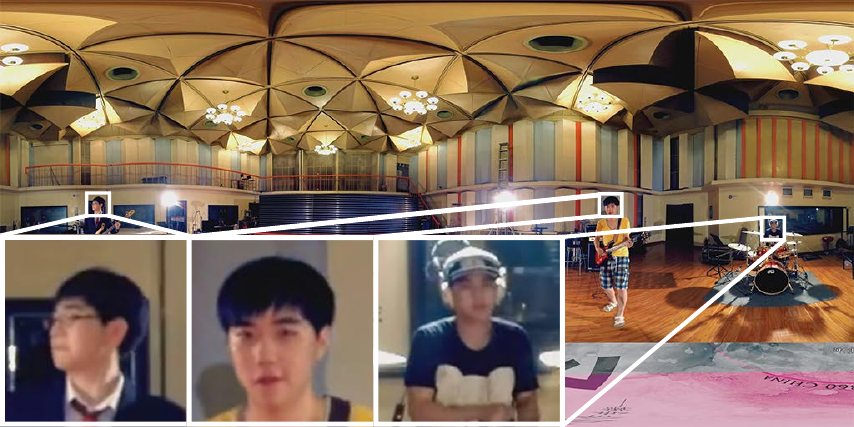}
    }\hspace{-0.5em}
    \subfigure[QP=37]{
    \label{fig:qp:37}
    \includegraphics[width=0.45\linewidth]{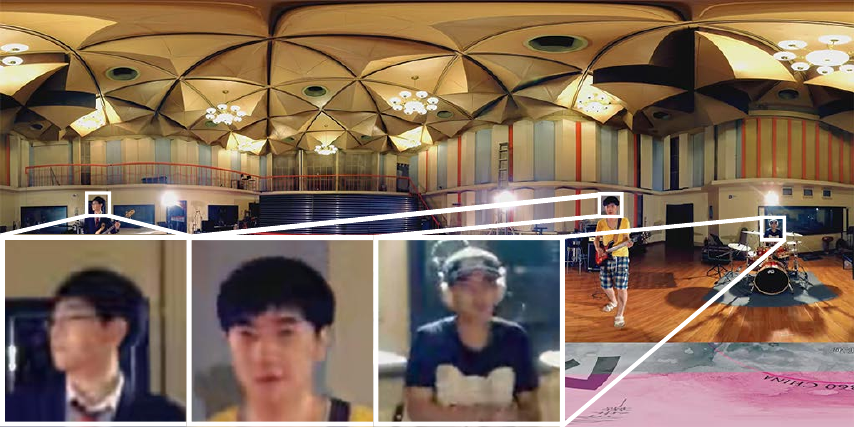}
    }\hspace{-0.5em}
    \subfigure[QP=42]{
    \label{fig:qp:42}
    \includegraphics[width=0.45\linewidth]{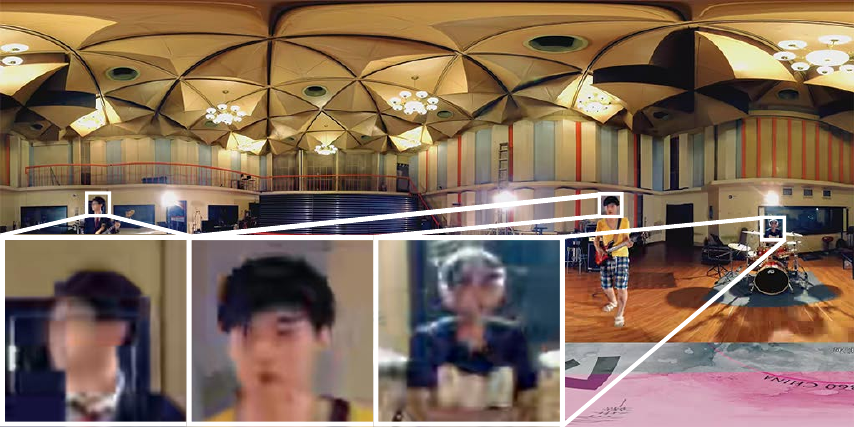}
    }
    \vspace{-1em}
    \caption{Illustration of a representative set of distorted frames (the 92th frame, \textit{Yanchanghui2}) compressed under different QP settings.}
    \vspace{-0.8em}
\label{fig:qp}
\end{figure}
\section{Experimental Results}\label{sec:experiment}
\begin{figure*}[!tb]
\centering
\subfigure[O-DMOS]{
  \label{fig:srcc:a}
  \includegraphics[width=0.23\linewidth]{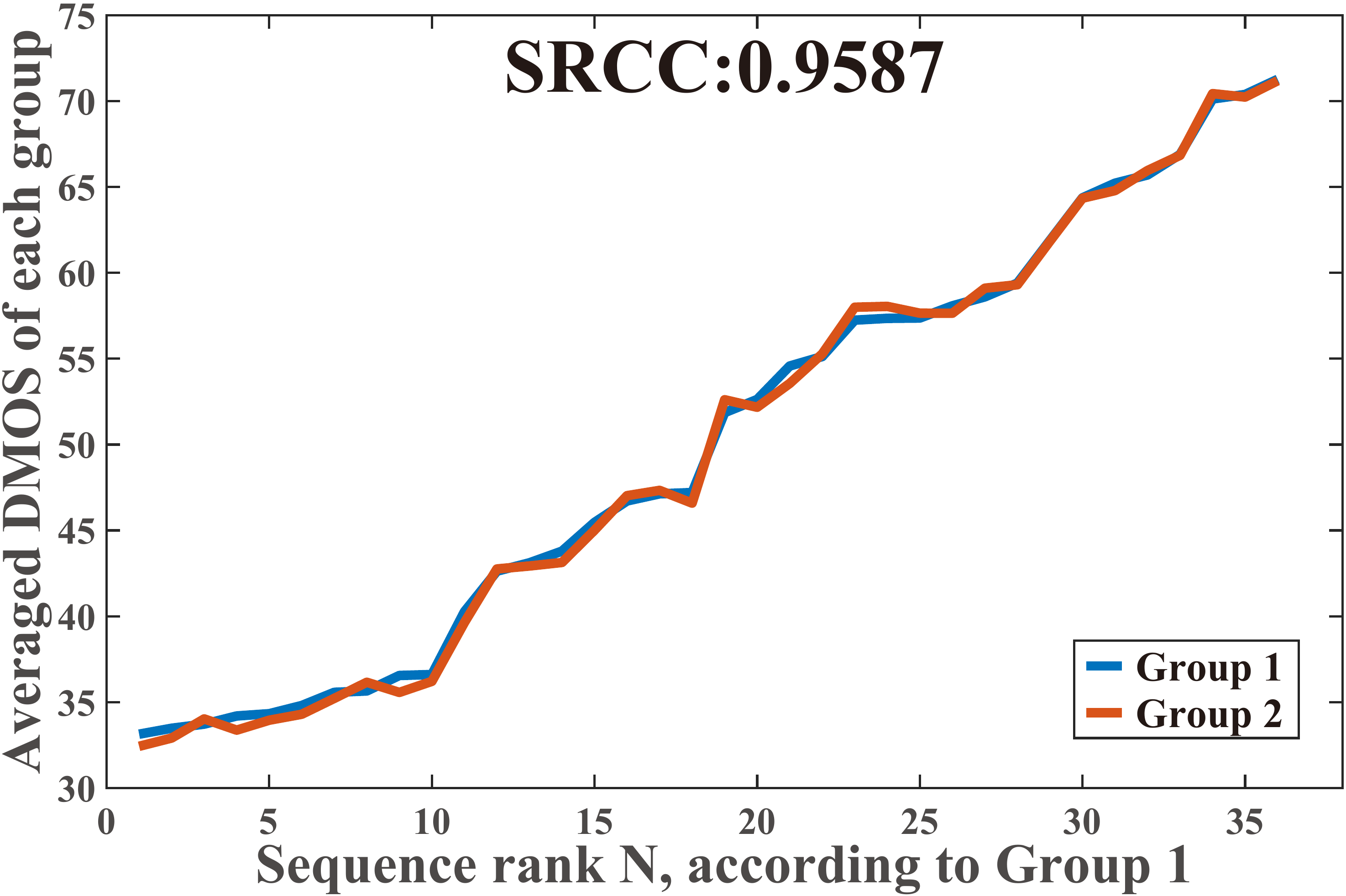}
}
\subfigure[V-DMOS (Front)]{
  \label{fig:srcc:b}
  \includegraphics[width=0.23\linewidth]{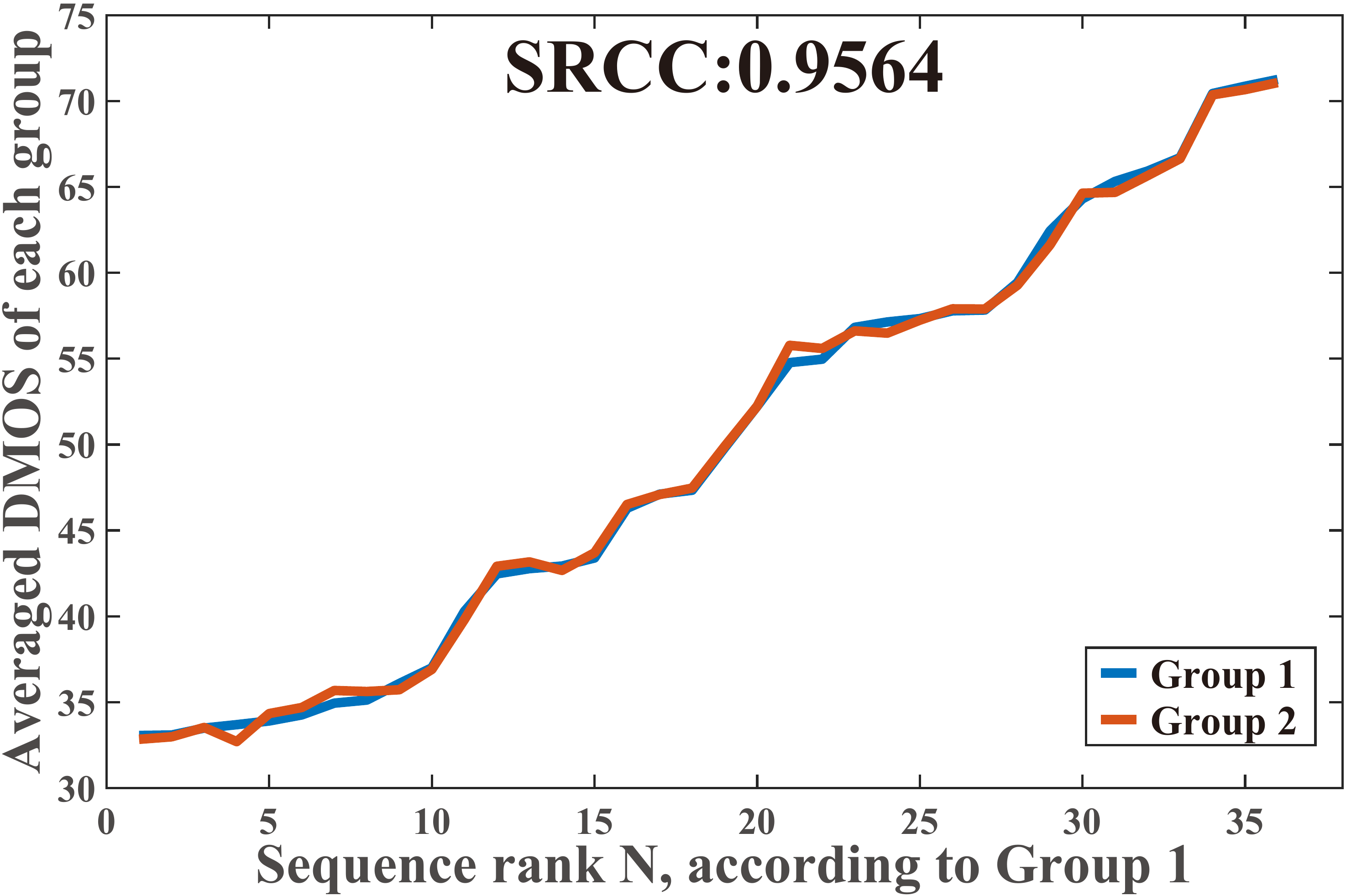}
}
\subfigure[V-DMOS (Left)]{
  \label{fig:srcc:c}
  \includegraphics[width=0.23\linewidth]{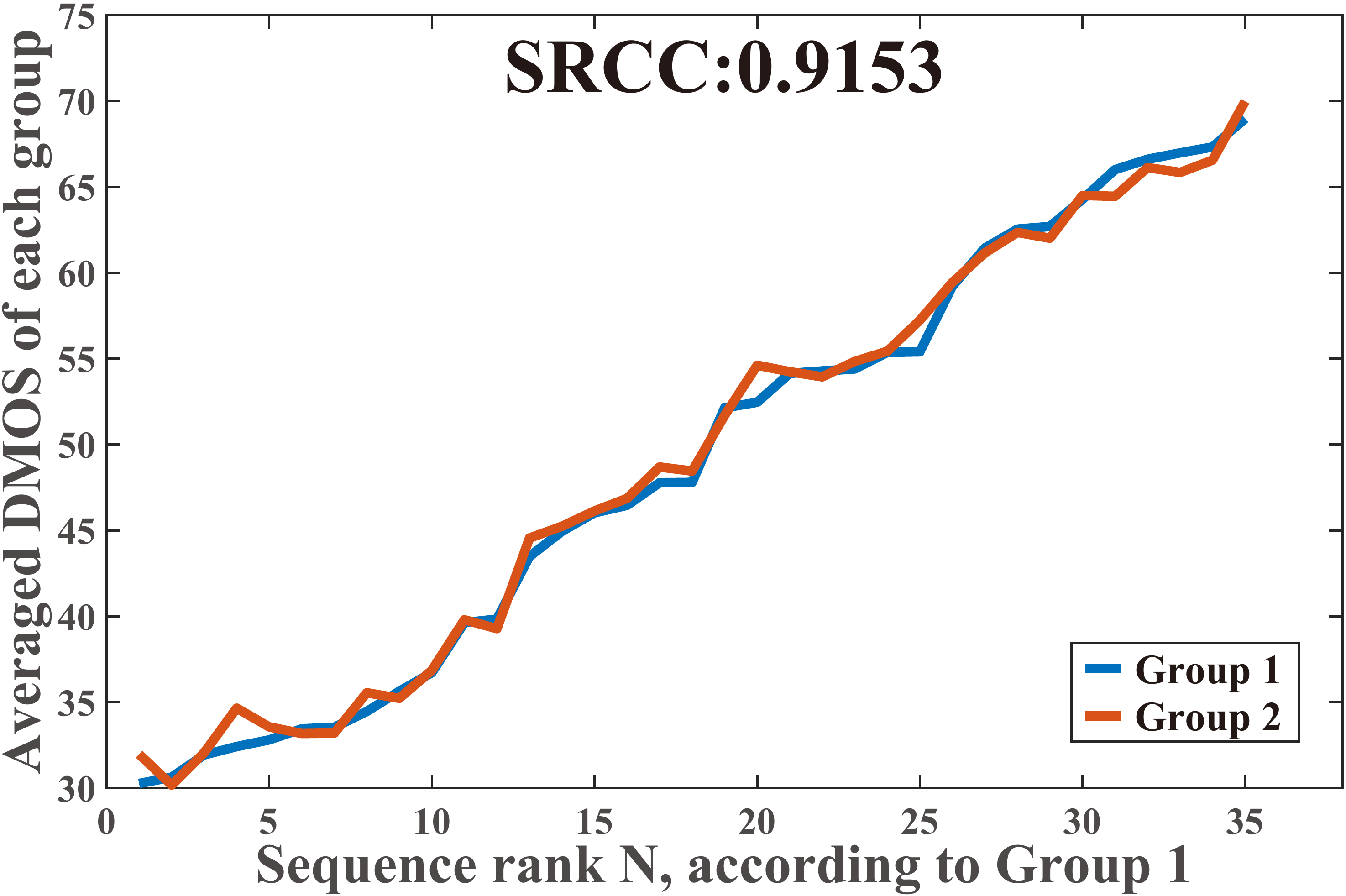}
}
\subfigure[V-DMOS (Right)]{
  \label{fig:srcc:d}
  \includegraphics[width=0.23\linewidth]{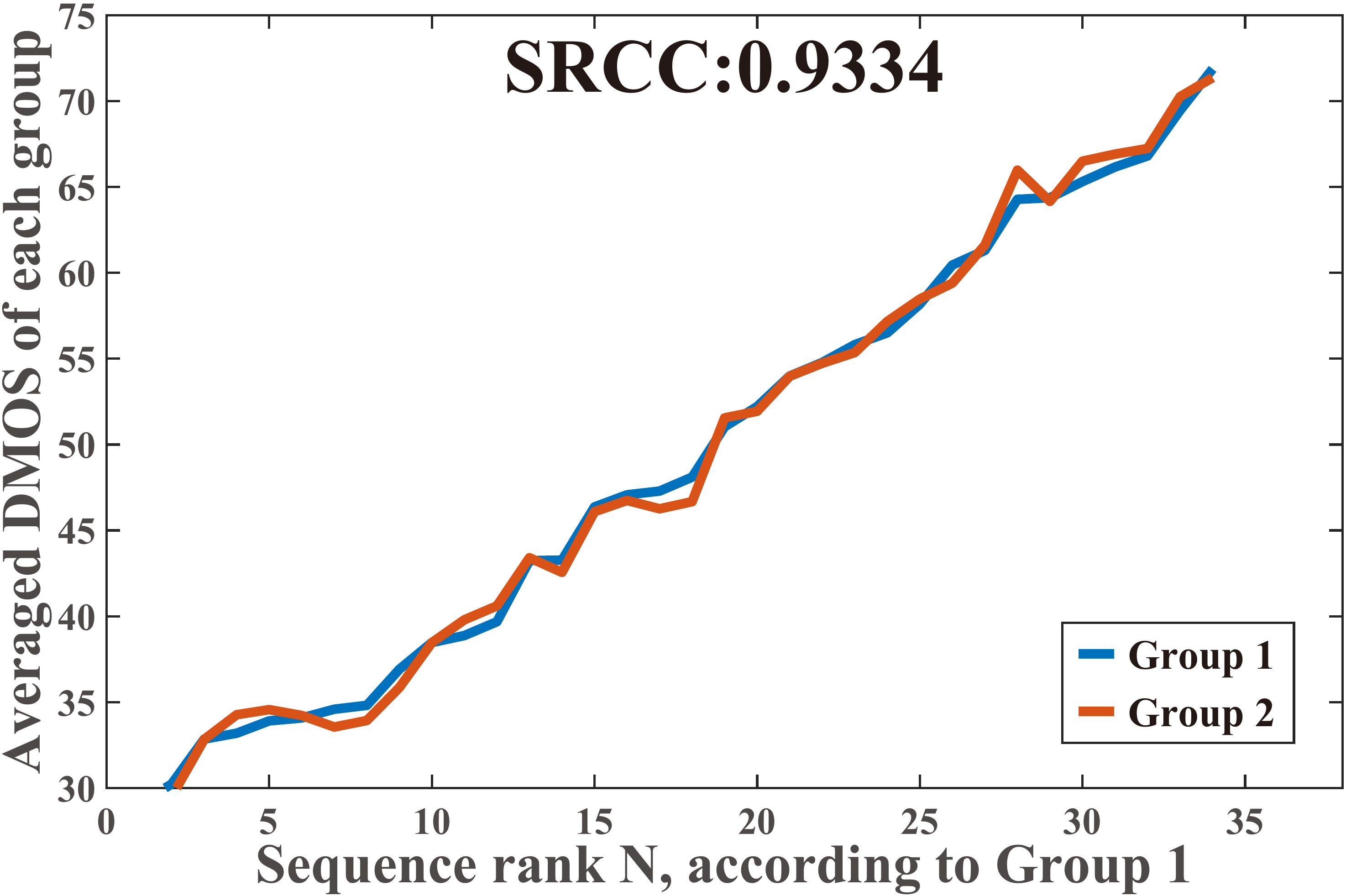}
}
\vspace{-1em}
\caption{Curves of the O-DMOS/V-DMOS values of each impaired sequence for two non-overlapping groups of subjects with equal size, in which the sequences are ranked in increasing order according to the O-DMOS/V-DMOS values of Group 1.}
\label{fig:srcc}
\vspace{-0.8em}
\end{figure*}

\subsection{Validation on our subjective VQA method}\label{sec:sub-experiment}
\textbf{Test benchmark and setting.} In this section, we validate the effectiveness of our subjective VQA method. First, all 12 uncompressed omnidirectional video sequences from \cite{vrsequences} (in YUV 4:2:0 format at resolution $4096\times2048$) are selected as the references. The duration of these sequences is 12 seconds with a frame rate of 25 frame per second (fps). Then, H.265/HEVC is used to compress these 12 sequences at 3 different bit-rates, under an equirectangular projection. For each sequence, the 3 bit-rates are set to be the actual bit-rates by quantization parameter (QP) = 27, 37 and 42. Figure \ref{fig:qp} shows a representative set of distorted frames compressed at these QP settings. Thus, there are 12 reference and 36 impaired sequences for the test in total.
Note that all test sequences are non-overlapping with 48 sequences of our viewing direction database introduced in Section \ref{sec:database} for fair comparison with other methods.

A total of 48 subjects participated in the subjective test for our VQA method (presented in Section~\ref{sec:method}).
The subjects are non-overlapping with those mentioned in Section \ref{sec:database}.
In the test, subjects were required to view and rate all sequences for raw subjective scores. Next, the O-DMOS and V-DMOS are calculated with the rated raw scores. Here, we simply set the threshold $f_0$ to be $1/6$ in the V-DMOS calculation, as there are 6 regions in our omnidirectional videos. It is worth mentioning that no subject was rejected in the calculation of the O-DMOS and V-DMOS values after using the subject rejection scheme from \cite{rec2012bt}. Finally, the values of the O-DMOS and V-DMOS obtained from the raw quality scores of the 48 subjects are reported in Table~\ref{tab:dmos}\footnote{Note that the O-DMOS is included in the V-DMOS as the first element and in bold in Table~\ref{tab:dmos}. The second to the seventh elements represent the DMOS scores of the front, left, back, right, top, and bottom regions, respectively.}.

\textbf{Evaluation on the effectiveness of our subjective VQA method.}
The effectiveness of our subjective VQA method is verified by evaluating the correlations between the O-DMOS/V-DMOS scores of different groups of subjects. Specifically, all 48 subjects are randomly and equally divided into two non-overlapping groups, Group 1 and Group 2, by 30 trials. Then, the O-DMOS/V-DMOS values are averaged over 30 trials, and the correlations of the averaged O-DMOS/V-DMOS values between two groups are evaluated as follows. Figure~\ref{fig:srcc} shows the curves of the ranked O-DMOS/V-DMOS values\footnote{Due to space limitations, we only show the values of the front, left and right regions for V-DMOS.} for all 36 impaired sequences obtained by Group 1, and the figure also presents the O-DMOS/V-DMOS values by Group 2 for the sequences ranked by the values for Group 1. We can see from this figure that the correlations between the two groups of O-DMOS/V-DMOS values are extremely high. We quantify such correlations using Spearman's rank correlation coefficient (SRCC), which is shown in Figure~\ref{fig:srcc}. The high SRCC values again indicate the agreement between the two groups for O-DMOS and V-DMOS. Since two randomly selected groups can be seen as the results from two subjective tests, the achieved agreement over different subjective tests implies that our method is effective in assessing subjective quality of omnidirectional video coding.

\textbf{Performance analysis of our subjective VQA method.}
\label{sec:perfomance}
It is necessary to ascertain the minimum number of subjects required for our subjective VQA method. To this end, we measure the SRCC of the O-DMOS values between the two groups with different numbers of subjects. Accordingly, Figure~\ref{fig:num-sub} shows the SRCC with an increased number of subjects in Groups 1 and 2, which is the averaged result over 30 trials. We can see that SRCC converges when the number of subjects is more than 20. Thus, we recommend 20 as the minimum number of subjects for our VQA method.

It is also interesting to investigate the relationship between the O-DMOS and V-DMOS values of different regions. Table~\ref{tab:OVSRCC} shows the SRCC results between the O-DMOS and V-DMOS values of different regions, which are calculated from all 48 subjects. It is clear that the V-DMOS values of the front, left and right regions have strong correlation with the O-DMOS values. In contrast, the V-DMOS values of the back and bottom regions are generally correlated with the O-DMOS values. However, the SRCC result for the V-DMOS values of the top region is rather small. It is because the V-DMOS values of the top region are determined by only few subjects, since most subjects pay no attention to the top region. In general, the correlations between O-DMOS and V-DMOS of different regions agree with \textit{Finding 2}, verifying the effectiveness of the proposed V-DMOS metric.
It can be concluded that the quality of different regions does not have equal contribution to the overall quality of the whole omnidirectional video. V-DMOS allows us to obtain the quality of different regions and their correlation with the overall quality of omnidirectional video. Then, we are able to adjust the optimization schemes, bit allocation and bandwidth allocation in coding and streaming for different regions of omnidirectional videos.
\begin{table}[!tb]
  \caption{SRCC between the O-DMOS and V-DMOS scores of different regions.} \label{tab:OVSRCC}
\vspace{-1.5em}
\begin{center}
\begin{tabular}{*{7}{|c}|}
  \hline
  \textbf{Region} & Front & Left & Right & Back & Top & Bottom \\
  \hline
  \textbf{SRCC} & 0.9972 & 0.9794 & 0.9750 & 0.8844 & -0.0857 & 0.8487 \\
  \hline
\end{tabular}
\end{center}
\vspace{-2.5em}
\end{table}
\begin{figure}[!tb]
\centering
  \includegraphics[width=0.9\linewidth]{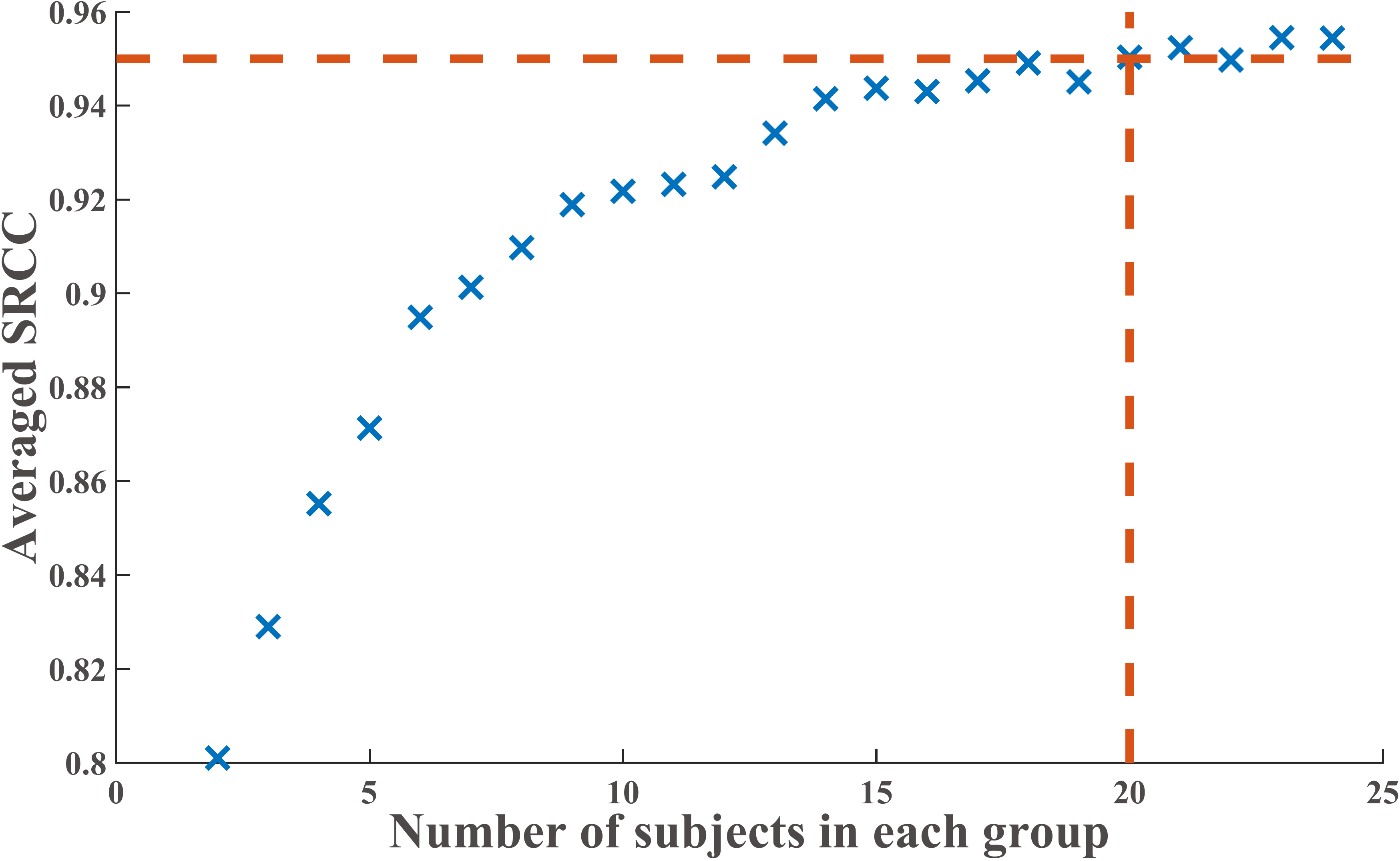}
\vspace{-1.2em}
\caption{SRCC of O-DMOS scores between two groups with increasing numbers of subjects in both groups, representing averaged result over 30 trials.}
\label{fig:num-sub}
\vspace{-0.8em}
\end{figure}
\begin{figure*}[!tb]
\centering
\subfigure{
  \includegraphics[width=0.2\textwidth]{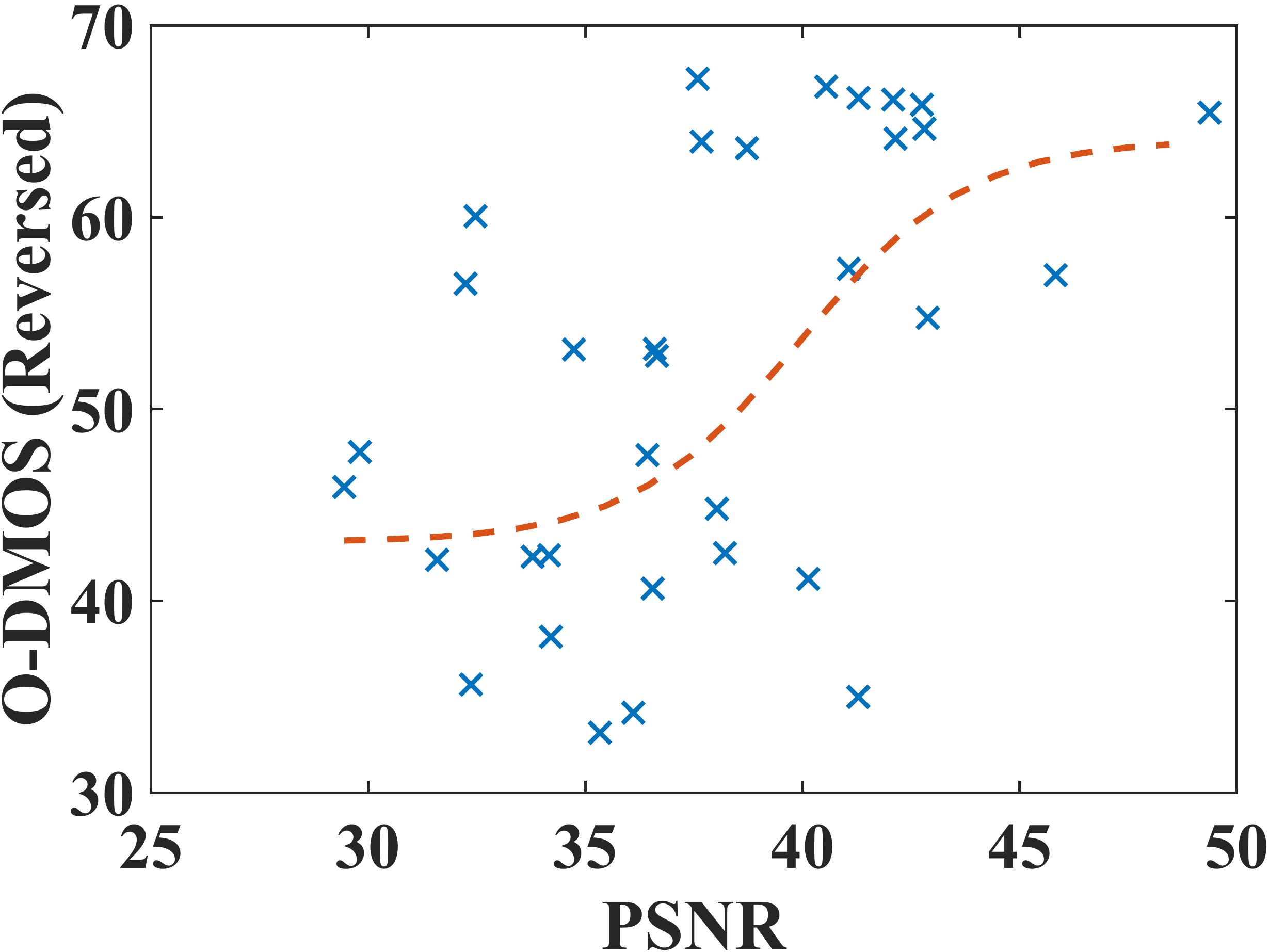}
}
\subfigure{
  \includegraphics[width=0.2\textwidth]{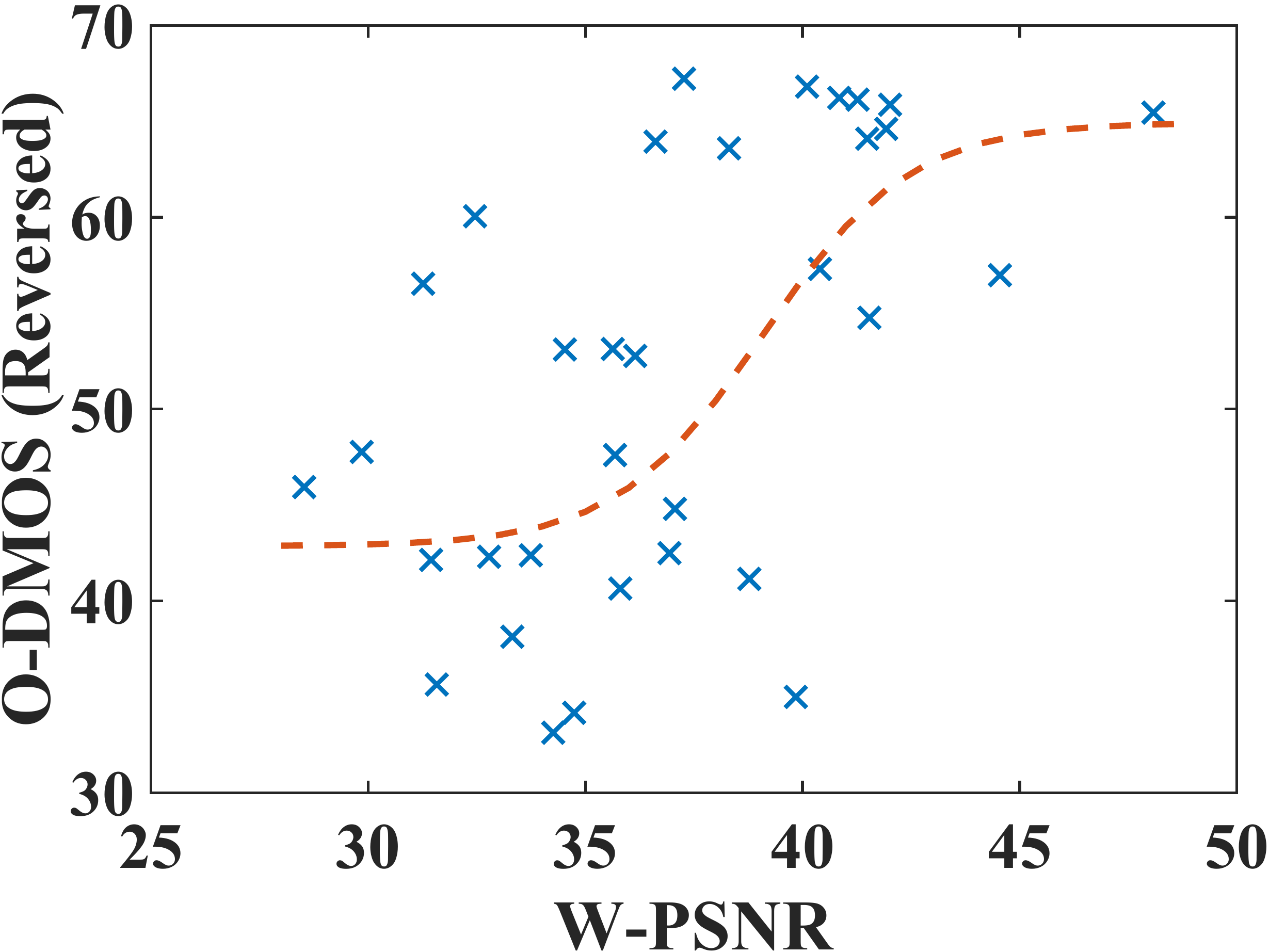}
}
\subfigure{
  \includegraphics[width=0.2\textwidth]{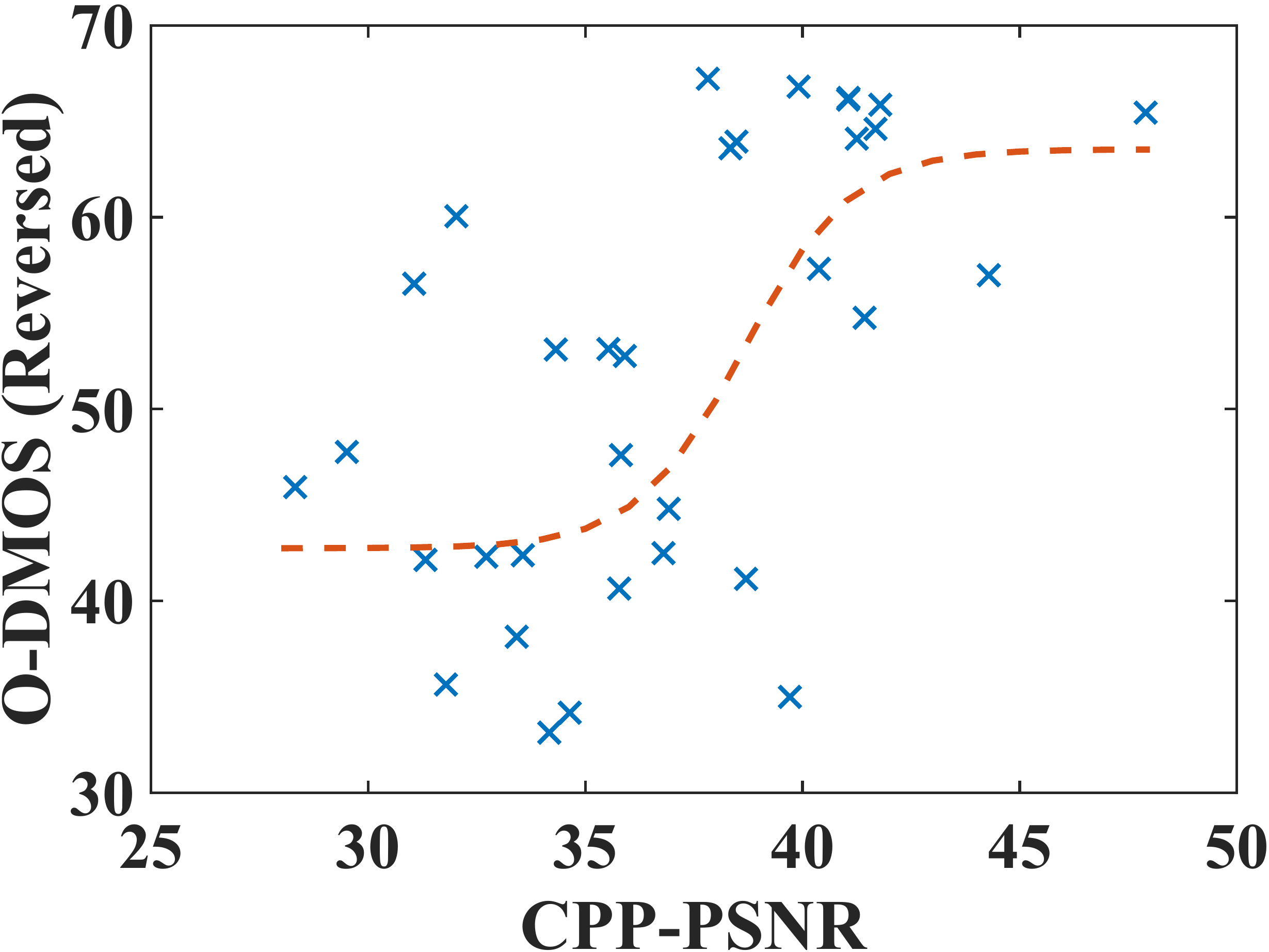}
}
\subfigure{
  \includegraphics[width=0.2\textwidth]{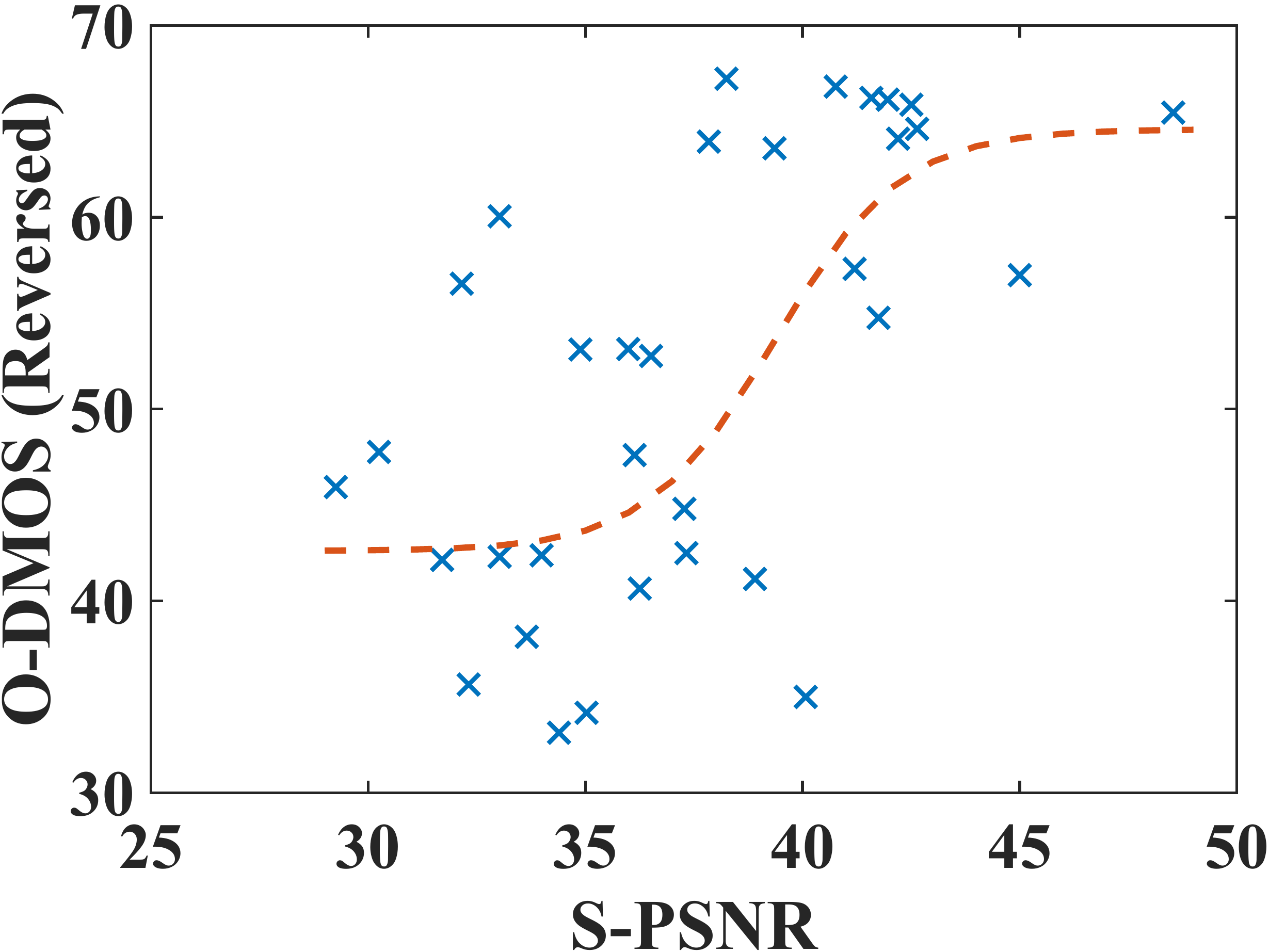}
}
\subfigure{
  \includegraphics[width=0.2\textwidth]{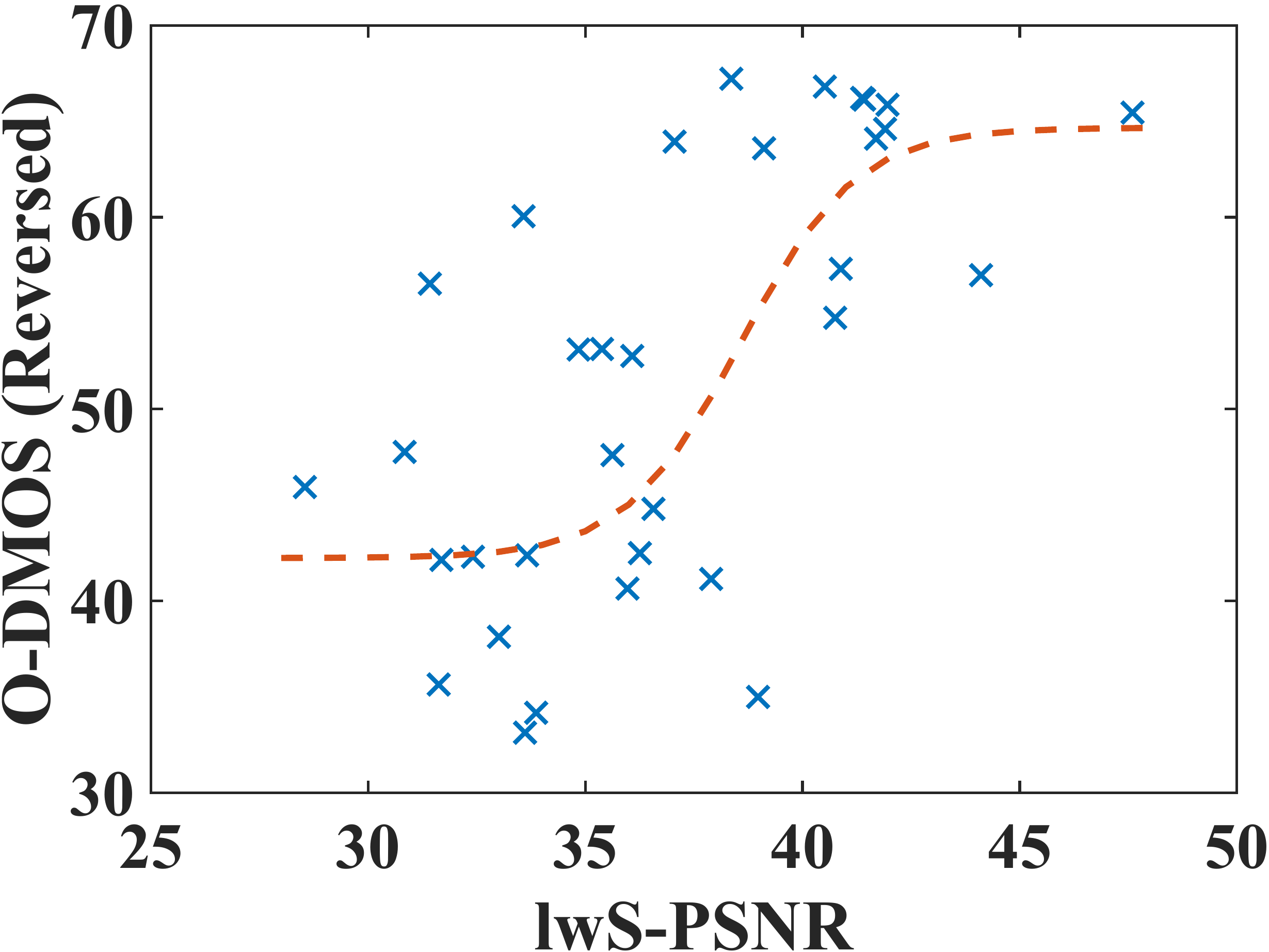}
}
\subfigure{
  \includegraphics[width=0.2\textwidth]{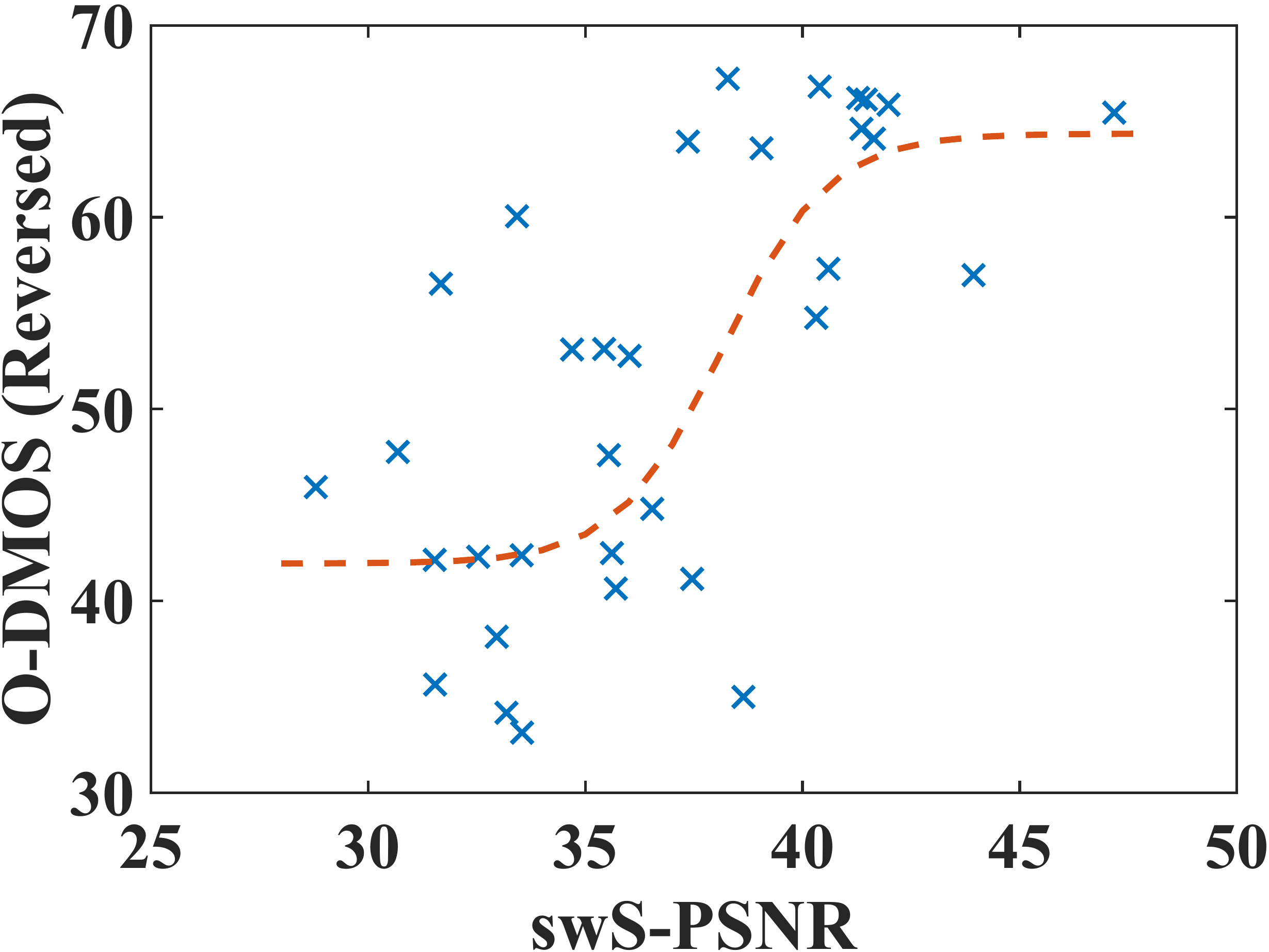}
}
\subfigure{
  \includegraphics[width=0.2\textwidth]{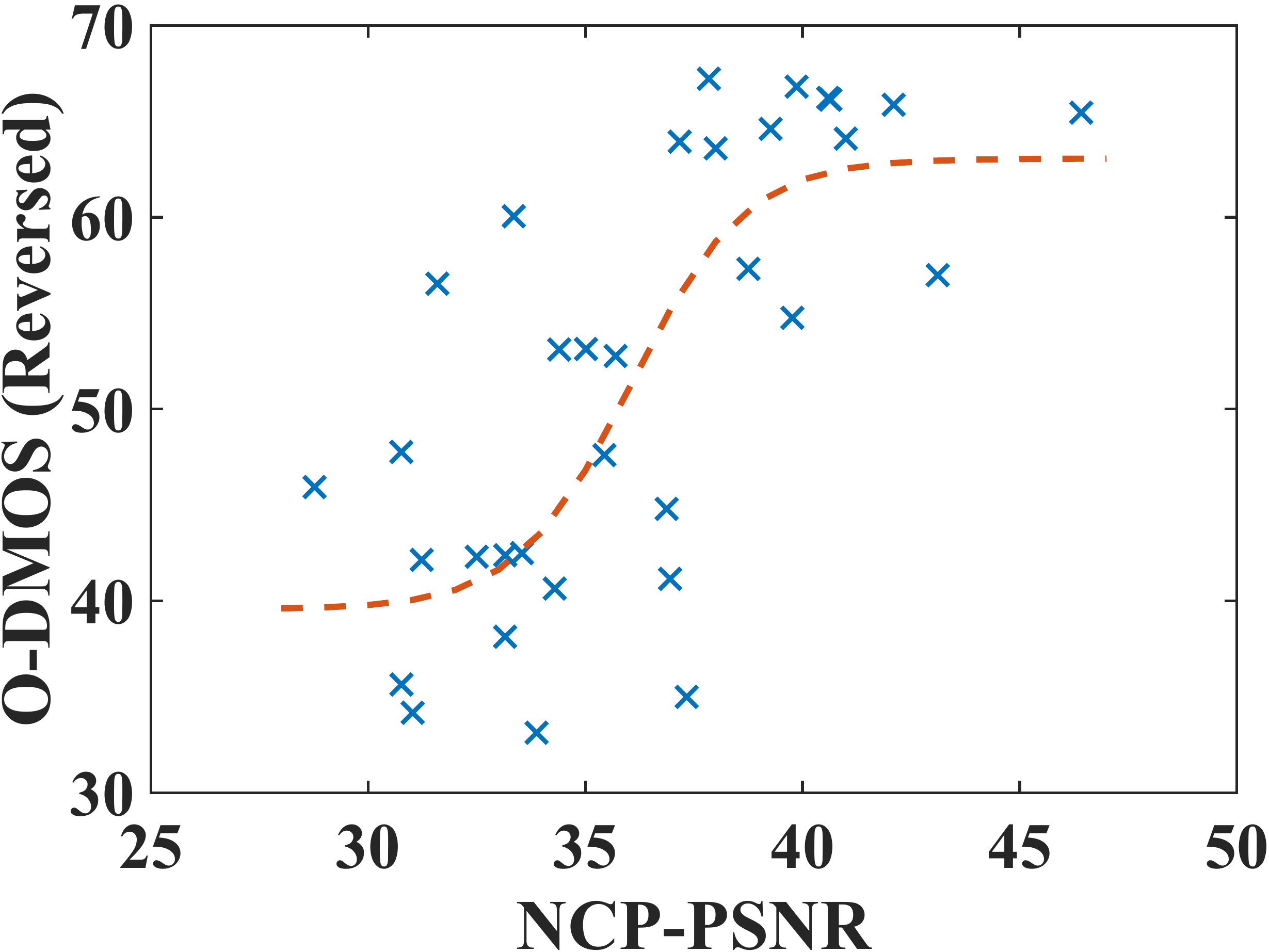}
}
\subfigure{
  \includegraphics[width=0.2\textwidth]{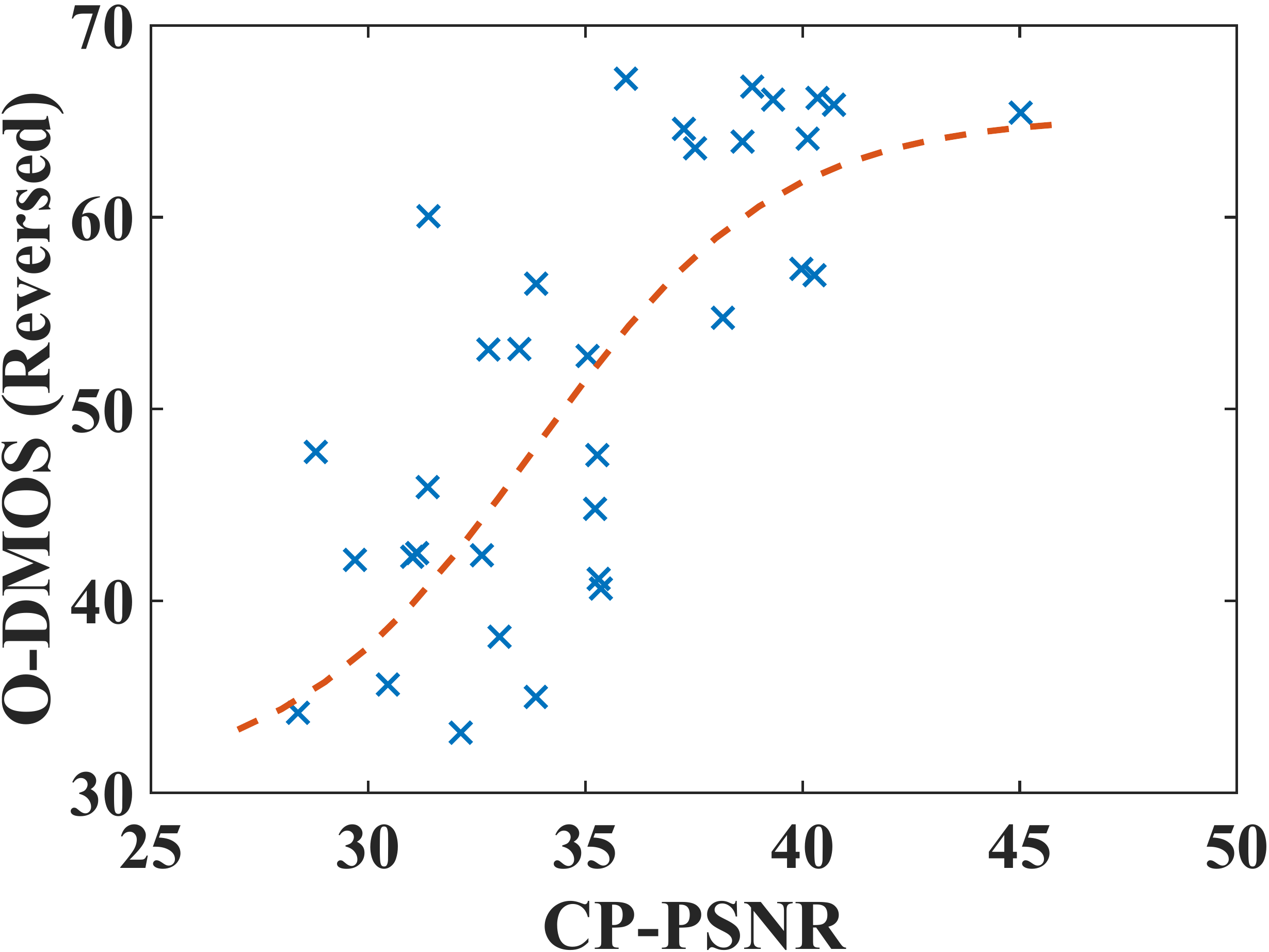}
}
\subfigure{
  \includegraphics[width=0.2\textwidth]{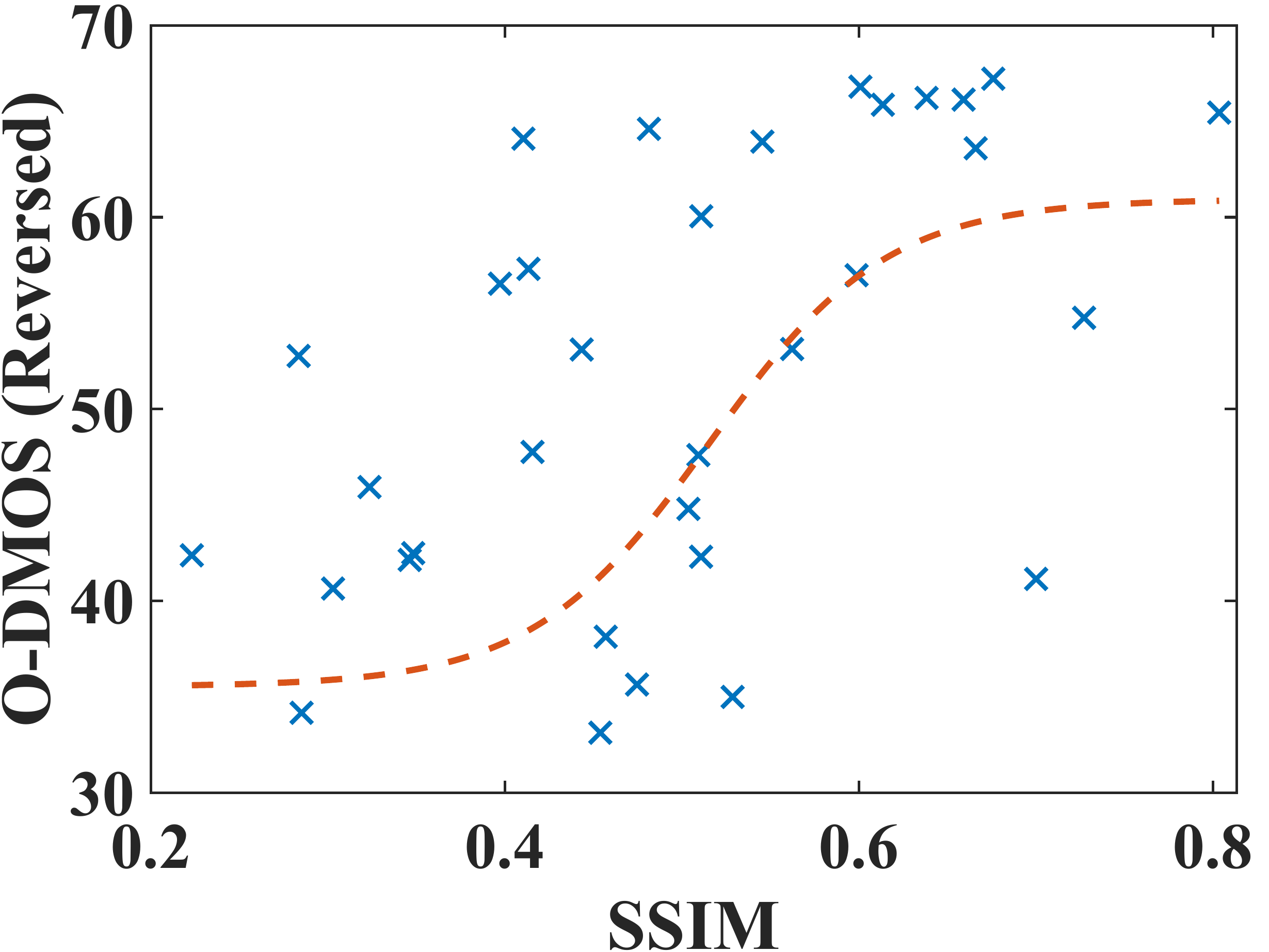}
}
\subfigure{
  \includegraphics[width=0.2\textwidth]{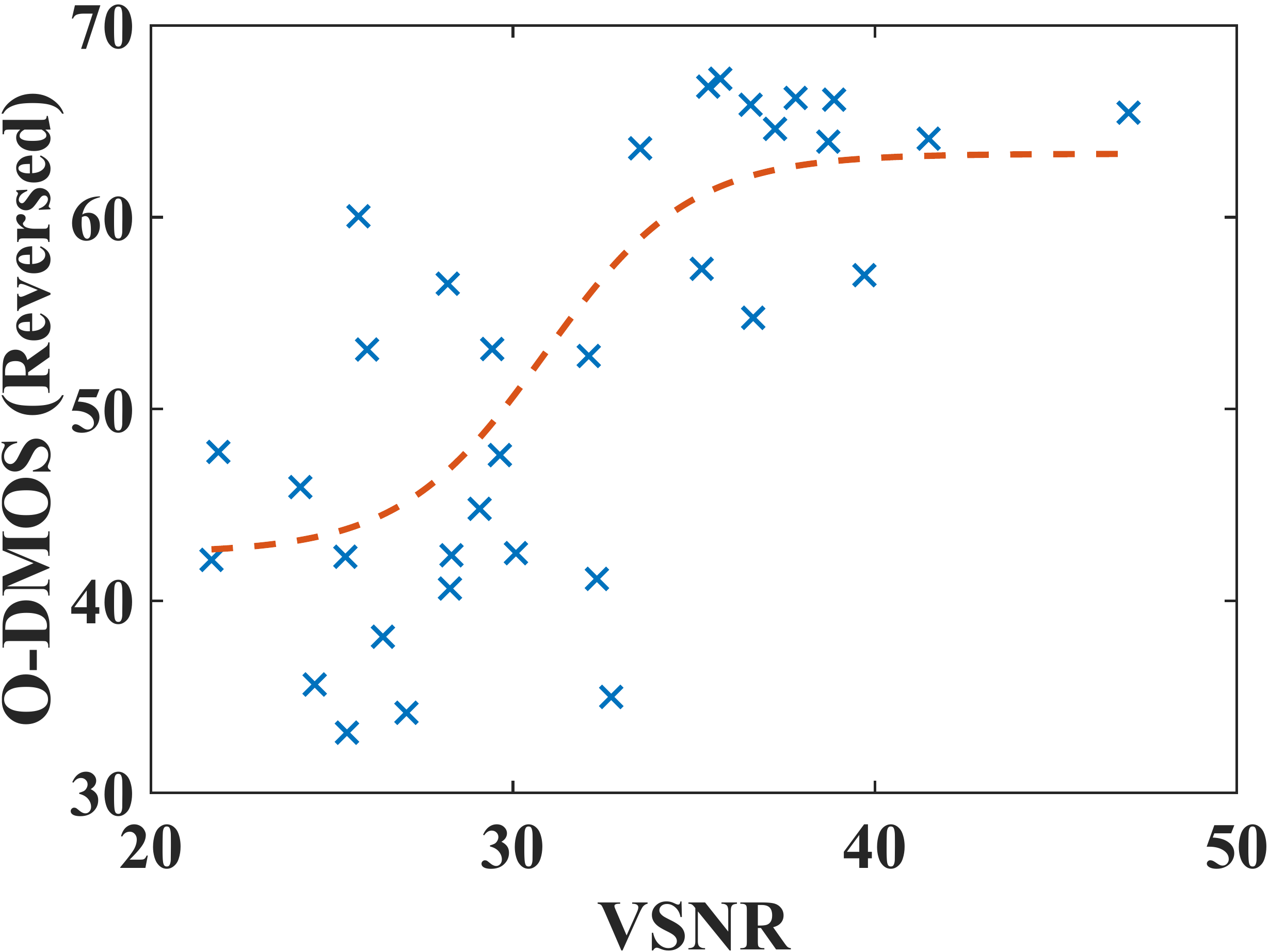}
}
\subfigure{
  \includegraphics[width=0.2\textwidth]{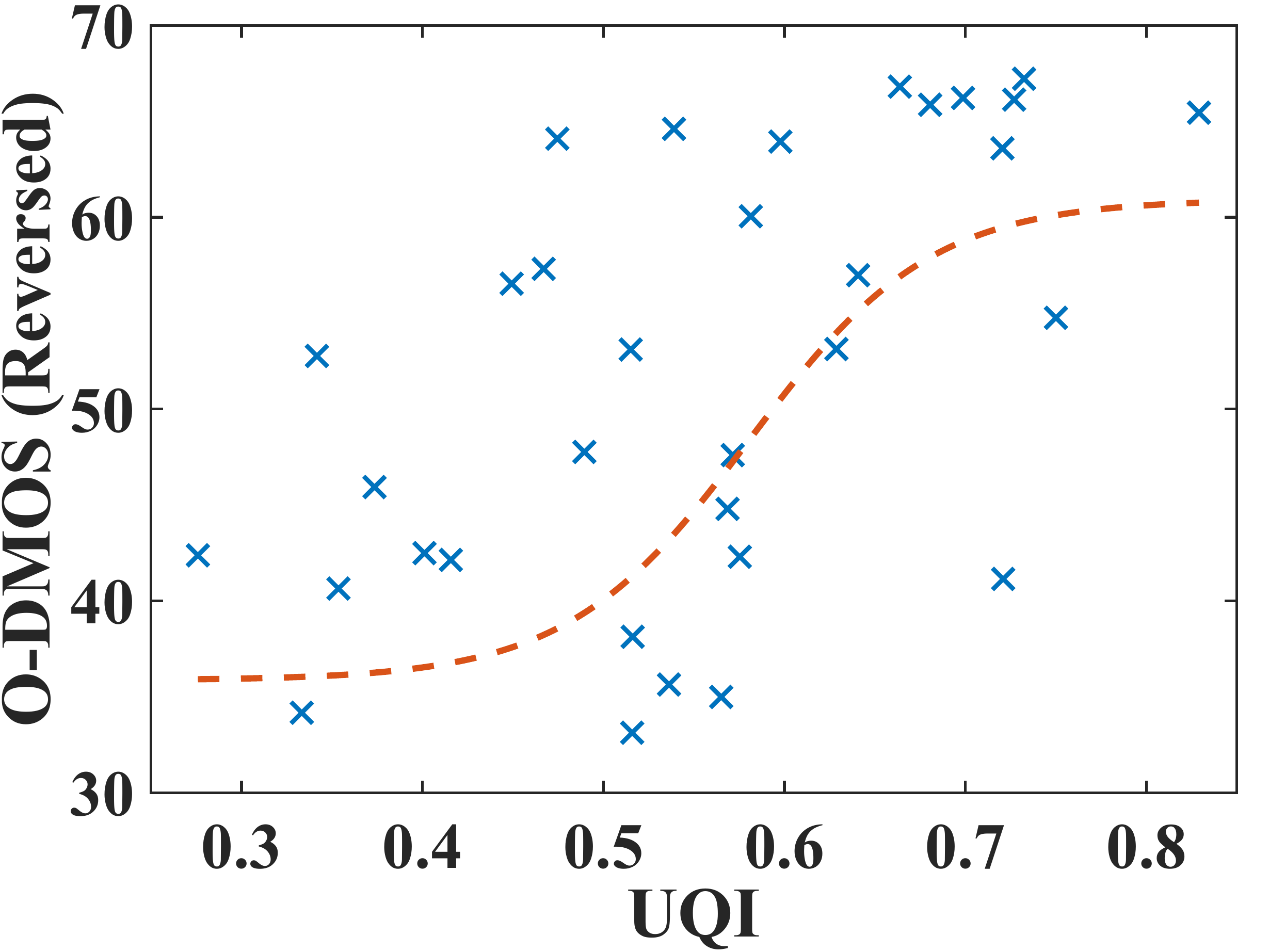}
}
\subfigure{
  \includegraphics[width=0.2\textwidth]{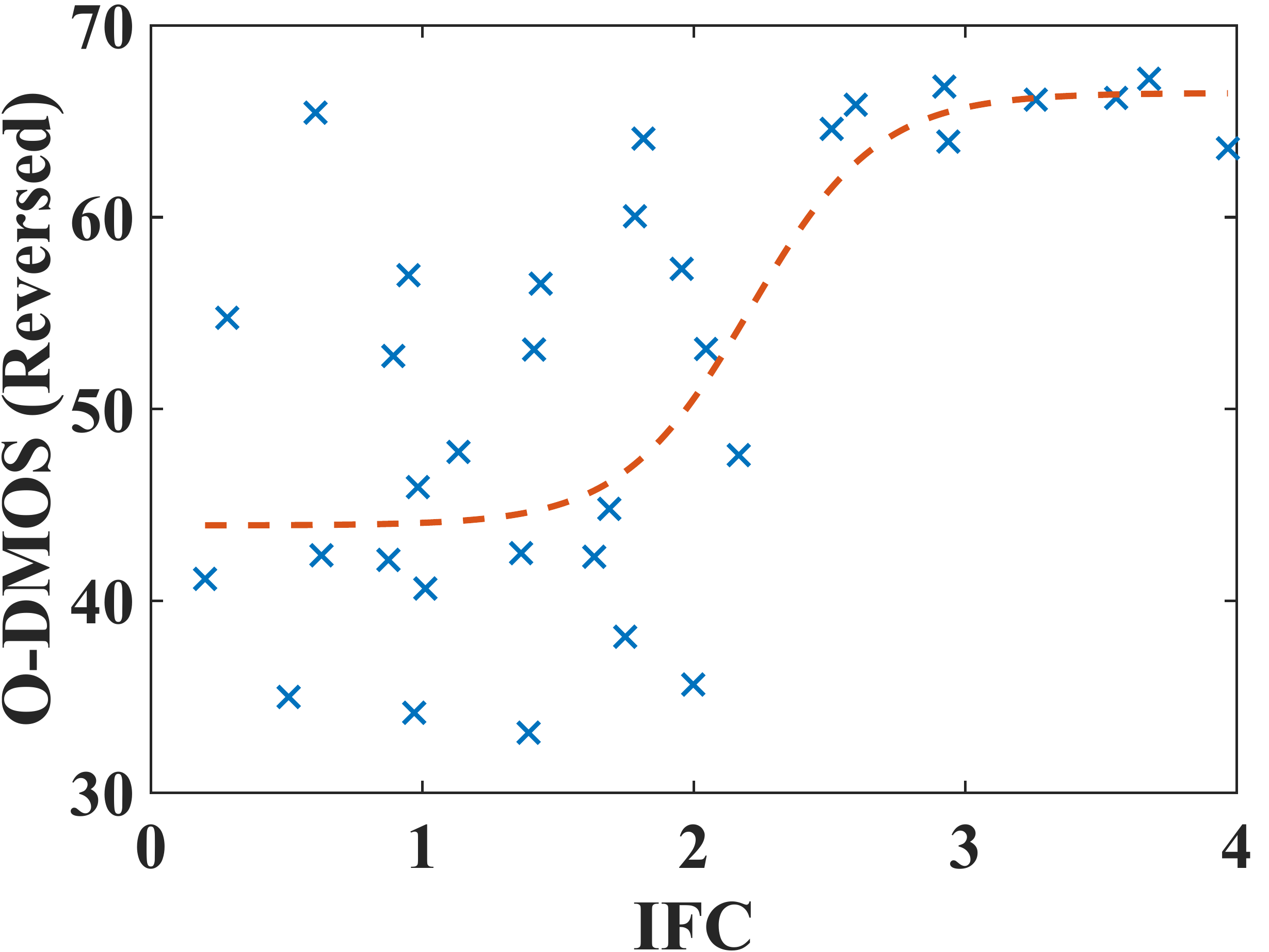}
}
\vspace{-1em}
\caption{Scatter plots of the objective VQA results versus the O-DMOS values for all 36 impaired sequences.}
\label{fig:obj-scatter}
\vspace{-0.8em}
\end{figure*}
\subsection{Validation on our objective VQA methods}
\textbf{Test benchmark and evaluation metrics.} The performance of our objective VQA methods is evaluated by measuring the agreement between subjective and objective quality. The performance evaluation is conducted on 36 impaired sequences of 12 uncompressed omnidirectional video sequences, as mentioned in Section \ref{sec:sub-experiment}.
Here, the subjective quality of those impaired sequences is the O-DMOS values of 48 subjects obtained in Section \ref{sec:sub-experiment}. For calculating CP-PSNR, all 48 omnidirectional video sequences from our viewing direction database presented in Section \ref{sec:database}, which does not overlap with any test sequence of this section, are used as the training data to learn the random forest model. All PSNR-related objective methods are calculated on the Y component and averaged over all frames for each impaired sequence.
Note that no parameter needs to be re-estimated on the test sequences for fair comparison.
Given the O-DMOS results, the performance of the objective VQA is measured with SRCC, Pearson correlation coefficient (PCC), Root-Mean-Square Error (RMSE) and Mean Absolute Error (MAE). SRCC measures the monotonicity of the objective quality with respect to subjective quality, while PCC quantifies the correlation coefficients between subjective and objective quality. In addition, RMSE and MAE measure the difference between the objective and subjective VQA results. Obviously, large-valued SRCC and PCC, or a small-valued RMSE and MAE, indicate a high degree of agreement between objective and subjective methods.
We follow \cite{seshadrinathan2010study} to apply a logistic function for fitting the objective VQA scores to the subjective O-DMOS scores in the performance evaluation for the objective VQA methods,
\begin{equation}
\small
\label{eq:logistic}
  Q'_j=\beta_2+\frac{\beta_1-\beta_2}{1+\mathrm{e}^{-\left(\frac{Q_j-\beta_3}{|\beta_4|}\right)}}\mbox{,}
\end{equation}
where $Q_j$ and $Q'_j$ are the original and fitted objective scores for sequence $j$, respectively. In \eqref{eq:logistic}, $\beta_1$, $\beta_2$, $\beta_3$ and $\beta_4$ are fitting parameters, initialized in the same way as \cite{seshadrinathan2010study}. The non-linear least squares optimization is performed to obtain the optimal parameters of $\beta_1$, $\beta_2$, $\beta_3$ and $\beta_4$. Then, SRCC, PCC, RMSE and MAE are calculated between the O-DMOS values and the fitted objective scores. Note that the O-DMOS values are reversed (i.e., subtracted from 100) for curve fitting with the logistic function.

\textbf{Comparison of scatter plots.}
Now, we compare our two objective VQA methods, NCP-PSNR and CP-PSNR, with traditional PSNR and five state-of-the-art methods. The five methods include S-PSNR, latitude-weighted S-PSNR (lwS-PSNR) and sphere-weighted S-PSNR (swS-PSNR), all of which are from \cite{yu2015framework}, as well as the latest W-PSNR and CPP-PSNR.
In addition to these PSNR based methods, four objective VQA methods for 2D video, SSIM, VSNR, UQI and IFC, are also calculated for comparison.
Figure \ref{fig:obj-scatter} shows the scatter plots of objective VQA results versus the O-DMOS results for all 36 impaired sequences along with the logistic fitting curves.
In general, intensive scatter points close to the fitting curve indicate high correlation of the objective VQA results with the subjective results, validating the effectiveness of the objective VQA method. It can be clearly seen from Figure~\ref{fig:obj-scatter} that the VQA results of our NCP-PSNR and CP-PSNR methods have a much higher correlation with the O-DMOS results, compared to other VQA methods. Therefore, we can conclude that both the NCP-PSNR and CP-PSNR perform far better than other methods.

\textbf{Comparison on quantification results.}
Furthermore, Table \ref{tab:obj-compare} reports the SRCC, PCC, RMSE and MAE between the fitted results of objective VQA and the subjective O-DMOS results, over all 36 impaired omnidirectional video sequences.
We can see from Table \ref{tab:obj-compare} that our NCP-PSNR and CP-PSNR improve the performance. The possible reasons of the improvements are as follows:
(1) As stated in Section VI-A, the distortion of pixels at different regions contribute unequally to the quality of omnidirectional video. The non-content-based weight map used in NCP-PSNR can reflect this inequality. Therefore, NCP-PSNR outperforms PSNR.
(2) Although other objective methods based on weight allocation also outperform PSNR, NCP-PSNR performs better than them. This indicates the non-content-based weight map is more effective.
(3) By predicting viewing directions with regard to video content, CP-PSNR can further emphasize the distortion in region of interest. This results in CP-PSNR further outperforming NCP-PSNR.
In addition, though based on PSNR, both NCP-PSNR and CP-PSNR perfrom better than the four non-PSNR-based methods. This further proves that the proposed NCP-VQA and CP-VQA methods are resultful in improving the performance of objective VQA methods of omnidirectional video.
\begin{table}[!tb]
 \centering
 \vspace{-0.5em}
  \caption{Comparison of the performance and the complexity of objective VQA methods.} \label{tab:obj-compare}
  \vspace{-1em}
  \resizebox{\linewidth}{!}{
  \begin{tabular}{|c|c|c|c|c|c||c|}
    \hline
    \multicolumn{2}{|c|}{Methods} & SRCC & PCC & RMSE & MAE & Time (second) \\
    \hline
    \multirow{8}{*}{PSNR based} & PSNR & 0.512 & 0.541 & 10.415 & 8.623 & 0.011 \\
    \cline{2-7}  & W-PSNR \cite{zakharchenko2016quality}& 0.556 & 0.596 & 9.938 & 8.097 & 0.025 \\
    \cline{2-7}  & CPP-PSNR \cite{zakharchenko2016quality}& 0.575 & 0.632 & 9.592 & 7.782 & \textit{3.857*} \\
    \cline{2-7}  & S-PSNR \cite{yu2015framework}& 0.589 & 0.639 & 9.518 & 7.692 & \textit{0.445*} \\
    \cline{2-7}  & lwS-PSNR \cite{yu2015framework}& 0.618 & 0.684 & 9.028 & 7.213 & \textit{0.462*} \\
    \cline{2-7}  & swS-PSNR \cite{yu2015framework}& 0.637 & 0.707 & 8.752 & 6.934 & \textit{0.445*} \\
    \cline{2-7}  & NCP-PSNR (our)& \textbf{0.702} & \textbf{0.725} & \textbf{8.539} & \textbf{6.770} & 0.025 \\
    \cline{2-7}  & CP-PSNR (our)& \textbf{0.751} & \textbf{0.764} & \textbf{7.991} & \textbf{6.657} & 2.405 \\
    \hline
    \multirow{6}{*}{Others} & SSIM \cite{wang2004image}& 0.547 & 0.562 & 10.391 & 8.684 & 0.585 \\
    \cline{2-7}  & VSNR \cite{chandler2007vsnr}& 0.684 & 0.741 & 8.741 & 6.888 & 1.431 \\
    \cline{2-7}  & UQI \cite{wang2002universal}& 0.586 & 0.594 & 10.774 & 8.943 & 0.428 \\
    \cline{2-7}  & IFC \cite{sheikh2005information}& 0.61 & 0.682 & 9.054 & 7.038 & 10.447 \\
    \cline{2-7}  & NCP-SSIM (our)& \textbf{0.802} & \textbf{0.799} & \textbf{7.443} & \textbf{5.984} & 0.599 \\
    \cline{2-7}  & CP-SSIM (our)& \textbf{0.815} & \textbf{0.807} & \textbf{7.307} & \textbf{5.977} & 2.980 \\
    \hline
    \multicolumn{7}{l}{*These methods are implemented in C++, while others are implemented in MATLAB.}
  \end{tabular}%
  }
\vspace{-1.5em}
\end{table}

\textbf{Statistical significance analysis}. We follow \cite{seshadrinathan2010study} to implement F-test on the residuals between objective VQA scores and DMOS values. Refer to the supporting document for the results of F-test. The results show that at 90\% significance level, both NCP-PSNR and CP-PSNR are superior to PSNR, indicating the significant improvement in performance of the proposed methods. Additionally, our CP-PSNR method is even superior to SSIM, UQI and W-PSNR. This further verifies the advantage of CP-PSNR.

\textbf{Complexity analysis}. For complexity comparison, we test the runtime of the objective VQA methods. The experiment is run on a computer with Intel\textregistered{} Core\texttrademark{} i7-8700 CPU. Each method is run with single thread on an omnidirectional video with resolution of $4096\times2048$.
Table \ref{tab:obj-compare} reports the computational time per frame in seconds of each method. Most of these methods are implemented in MATLAB, except CPP-PSNR and the three variations of S-PSNR, which are implemented in C++. Since C++ runs at least twice faster than MATLAB \cite{andrews2012computation}, the results of these four methods are highlighted in italic in Table \ref{tab:obj-compare}.
In Table \ref{tab:obj-compare}, the runtime of NCP-PSNR equals to that of W-PSNR, since the weight maps are generated offline and the additional time is only spent on applying the weight maps in calculation. In general, NCP-PSNR runs rather fast but brings better performance than most of the methods. Although CP-PSNR reaches the best performance among these methods at the cost of high complexity, it still runs faster than CPP-PSNR and IFC. In conclusion, our methods achieves the best performance with modest computational complexity increment.

\textbf{Extension to other methods}. Our VQA methods can be easily extended to other methods, including SSIM. Here, we extend our NCP-VQA and CP-VQA methods to SSIM, producing NCP-SSIM and CP-SSIM. NCP-SSIM and CP-SSIM can be obtained by
\begin{eqnarray}
\small
\label{eq:NCP-SSIM}
\text{NCP-SSIM} = \sum_{s,t}{{m_{\mathrm{SSIM}}(s,t) \cdot \tilde{w}(s,t)}}\mbox{,}\\
\text{CP-SSIM} = \sum_{s,t}{{m_{\mathrm{SSIM}}(s,t) \cdot \widetilde{w'}(s,t)}}\mbox{,}
\end{eqnarray}
where $m_{\mathrm{SSIM}}$ is the SSIM map with local SSIM value for pixel $(s,t)$; $\tilde{w}$ and $\widetilde{w'}$ are the normalized non-content-based and content-based weight maps proposed in \eqref{eq:wtilde} and \eqref{eq:wtilde2}. Table \ref{tab:obj-compare} shows the performance and complexity of these two methods.
We can see that on the basis of SSIM, the performance of our NCP-SSIM and CP-SSIM methods is rather high in terms of the four metrics, better than that of NCP-PSNR and CP-PSNR. However,  the cost is additional computational complexity.
In conclusion, our NCP-VQA and CP-VQA methods can be easily extended and with improved performance.
\section{Conclusion}
In this paper, we have proposed both subjective and objective VQA methods for evaluating the quality degradation of impaired omnidirectional video.
In contrast with the conventional VQA methods, human viewing directions were investigated and then taken into account in our VQA methods.
Specifically, we conducted an experiment to present a new database, which contains the viewing directions from 40 subjects on viewing 48 omnidirectional video sequences. Next, we found from our database that subjects consistently prefer looking at the center of front region of omnidirectional video, but there still exists dependency on video content for viewing directions.
In light of our findings, we proposed two subjective VQA metrics, O-DMOS and V-DMOS, measuring the overall and regional quality reduction of impaired omnidirectional video, repsectively.
In addition, we proposed two objective methods, NCP-PSNR and CP-PSNR, for assessing the quality loss of compressed omnidirectional videos. In NCP-PSNR, the quality loss is weighed according to statistical results on the preference for the center of the front region, while CP-PSNR imposes quality loss using weights with respect to possible viewing directions predicted upon the video content.
Finally, our experimental results validate the effectiveness of our subjective and objective VQA methods.
Between the two proposed methods, NCP-PSNR can be calculated at a fast speed, while CP-PSNR can achieve better performance with the extra runtime.

There are two promising directions for future work.
First, in addition to FoV, there may be some other characteristics of the HVS benefiting for VQA of omnidirectional video. For example, there are some high level features, such as local motion and objectness, can also be incorporated in predicting viewing direction to improve the CP-VQA method.
This is an promising future work.
Second, future work may apply our VQA method in optimizing the encoder of omnidirectional video coding. For example, NCP-PSNR or CP-PSNR can be maximized in the bit allocation when encoding omnidirectional video.
\bibliographystyle{IEEEtran}
\bibliography{trasfer}
\end{document}